\documentclass[%
preprint,
 amsmath,amssymb,
aps,
prf,
]{revtex4-2}

\usepackage{graphicx}
\usepackage{dcolumn}
\usepackage{bm}
\usepackage{newtxtext}
\usepackage{newtxmath}
\usepackage{hyperref}
\usepackage{subcaption}
\usepackage{amsmath}
\usepackage{amsfonts}

\usepackage{dashrule} 
\usepackage{xcolor} 

\DeclareMathAlphabet\mathsfbi            {OT1}{cmss}{m}{sl}

\definecolor{mplblue}{RGB}{0,114,189}
\definecolor{mplred}{RGB}{213,94,0}
\definecolor{mplgreen}{RGB}{86,177,76}
\definecolor{black}{RGB}{0,0,0}

\begin{document}


\title{Effect of streaks on hypersonic boundary layer linear instability}

\author{Clément Caillaud}
\email{clement.caillaud@cea.fr}
\affiliation{CEA-CESTA, Le Barp, France}
\altaffiliation[Previously at ]{Institut Pprime, ISAE-ENSMA, Chasseneuil du Poitou, France}
\author{Guillaume Lehnasch}
\author{Eduardo Martini}
\altaffiliation[Previously at ]{CEA-CESTA, Le Barp, France}
\affiliation{Institut Pprime, ISAE-ENSMA, Chasseneuil du Poitou, France
}
\author{Peter Jordan}
\affiliation{%
 Institut Pprime, CNRS, Chasseneuil du Poitou, France
}%

\date{\today}

\begin{abstract}

Hypersonic boundary layers exhibit diverse transition pathways, influenced by various flow conditions and environments. Non-modal mechanisms, such as the lift-up effect, are recognized as pivotal contributors to transition, particularly over complex geometries like rough walls or blunt forebodies. The lift-up mechanism generates streaks that alter the baseflow, resulting in a modulated boundary-layer such that new unstable modes may arise. In this numerical study, we generate streaky baseflows by introducing optimal disturbances at the inlet of the computational domain. These baseflows serve as the foundation for investigating the linear dynamics supported by a streaky boundary layer evolving over an adiabatic flat plate at Mach number $M_\infty=6.0$. Our analysis involves linearized direct numerical simulations using white noise forcing, and locally parallel stability analysis combined with dedicated post-processing that we use to characterise the observed linear dynamics. The data processing involves an efficient SPOD extension based on Floquet theory in wavenumber space. The results reveal rich linear dynamics, quite different from the baseline case without streaks. Notably, low-frequency first-mode instabilities and streak-supported instabilities dominate the growth mechanisms. The results reveal and clarify a stabilisation of the second Mack mode by streaks of low to medium amplitude; and a destabilisation of the first mode. The study illustrates the rich linear dynamics that can arise due to the growth of streaks and how multiple paths are available for the transition to turbulence.

\end{abstract}

\maketitle

\newpage
\section{Hypersonic boundary layers and streaks}
\label{sec:intro}


Understanding the transition to turbulence on hypersonic bodies is critical from a design perspective \citep{schneiderHypersonicLaminarTurbulent2004}. Transition is underpinned by linear growth mechanisms that depend on shape of the body, the flow conditions and the base flow that results. 

Among the mechanisms underpinning the boundary layer state that supports the linear growth mechanisms responsible for transition, 
the lift-up effect effect leads to the appearance of steady streaks \citep{schmidNonmodalStabilityTheory2007, sippDynamicsControlGlobal2010, ellingsenStabilityLinearFlow1975a, landahlNoteAlgebraicInstability1980}. Lift-up is known to play a key role in hypersonic flows. For instance, it has been observed on blunt geometries \citep{zanchettaKineticHeatingTransition1996,paredesNosetipBluntnessEffects2018} ; in the wake of isolated or distributed roughness elements \citep{padillamonteroAnalysisStabilityFlatPlate2021,detullioLaminarTurbulentTransition2013} ; on cones \citep{laibleNumericalInvestigationHypersonic2011,haderDirectNumericalSimulations2019} ; from non-linear interaction in transitional flows \citep{frankoBreakdownMechanismsHeat2013} ; or at the reattachment of separation bubbles in corner flows \citep{caoUnsteadyEffectsHypersonic2021,dwivediObliqueTransitionHypersonic2021,lugrinTransitionScenarioHypersonic2020}. In all configurations, streaks are either present as a baseflow feature originating from the geometry or as a by-product of non-linear processes in the transition dynamics. Once present, they provide new possibilities for linear instability. It is therefore of interest to study the linear dynamics of streaky baseflows so as to explore, how the instability are impacted by streaks, and, in particular, how instability depends on streak amplitude. 

{
Since lift-up and streaks appear as a common feature of boundary-layer flows, it is worth recalling the dynamics observed for incompressible flows. In this regime, the growth of streaks and their effects have been thoroughly investigated. \cite{anderssonOptimalDisturbancesBypass1999,anderssonBreakdownBoundaryLayer2001} used local stability analysis and DNS to highlight the importance of the streaks for Tollmien-Schlichting (TS) waves. They showed that "sinuous" and "varicose" instabilities may arise.  Moreover, the effect of streak amplitude was studied and it was shown that new convectively unstable modes can arise when streak amplitudes are high. These streak-dependent instabilities were shown to have growth rates that exceed those of TS waves. As for the existence of absolute instabilities on saturated streaks, \citet{brandtConvectivelyUnstableNature2003} showed that even at very high amplitude saturated streaks remained stable and suggested the streaky baseflow to be a noise amplifier system only. 

A number of numerical and experimental studies have now confirmed that Tollmien-Schlichting waves are stabilized and even suppressed by low-to-moderate-amplitude streaks, and this has led to effective passive control strategies for such instabilities \citep{cossuStabilizationTollmienSchlichting2002, franssonExperimentalStudyStabilization2005}. Such results suggest streak production  as a promising candidate for stabilisation of high-speed compressible boundary layer as well.

Despite their importance in transition, effect of streaks on transition streaks have not been so comprehensively studied in hypersonic flows. As opposed to subsonic and supersonic flows, hypersonic boundary layers are known to support not one but two instability mechanisms: the first mode which is the continuation of TS waves in compressible flows and is shear driven. ; and the second mode, specific to hypersonic conditions, which comprises a trapped acoustic wave in the subsonic near-wall region \citep{mackBoundaryLayerLinearStability1984}. Both of these modes are known to be amplified in hypersonic,adiabatic boundary layers. With the particularity that, for cold isothermal boundary layers, the second mode is destabilised while the first-mode is getting stabilised \citep{fedorovHighSpeedBoundaryLayerInstability2011}. These two mechanisms are known to coexist with streaks in hypersonic flow conditions.  However, research on the stability properties of these modes in hypersonic streaky boundary layers is incomplete. Twenty years after the first computation of optimal transient growth  of streaks in compressible flows by \citet{hanifiTransientGrowthCompressible1996}. 
\citet{paredesTransientGrowthAnalysis2016} performed a characterisation of compressibility effects on optimal streaks using PSE for a supersonic flat plate boundary layer at $M_\infty=3.0$. The stability properties of oblique first-mode waves were found to be modified by the streaks by \cite{paredesInstabilityWaveStreak2017}. Stabilising effects were observed for waves having a spanwise wavelength greater than two streak wavelengths. On the contrary, shorter-wavelength first-modes were destabilised. Furhermore, at $M_\infty=3$, subharmonic first-mode waves were found to be the most amplified and led earlier to transition N-factor $N=5$ than the non-streaky boundary layer for moderate streak amplitude \citep{paredesTransitionDueStreamwise2016}. Another analysis was made in hypersonic conditions by \cite{paredesInstabilityWaveStreak2019} for a sharp-cone boundary layer with a cold isothermal wall at a free-stream Mach number $M_\infty=5.3$. The analysis focused only on second-mode waves which were shown to be stabilised by moderate-amplitude streaks. Specifically, planar second-mode waves were found to be stabilised both by increasing streak amplitude and varying streak spanwise wavelength. The latter parameter was shown to be the most important in achieving substantial transition delay for low to moderate streak amplitudes on the cone. On the other hand, no studies were performed for adiabatic wall conditions or first-mode waves growth in the hypersonic regime (i.e. $M_\infty>5$).

These studies show that streaks may, similar to what is observed in incompressible flow, lead to a stabilisation at high speed. However, the foregoing studies are not comprehensive in the sense that they focus each time on a selected first- or second-mode wave in supersonic streaky flow. We aim therefore, to complement these studies using a linearised global framework to explore the effects of streaks on linear growth mechanisms in general. Our objective is to characterise and classify the different kinds of linear growth mechanisms that may arise in streaky boundary layers, and to assess the effect of streak amplitude on these. This done by a first global linear analysis of streaky baseflows subjected to a broadband stochastic forcing, so as to obtain an unbiased view of the full range of linear dynamics that may arise. The global analysis is supported by 2D locally parallel stability calculations and an N-periodic SPOD formulation based on Floquet theory. Thanks to this combination of global, forced, linear simulations, locally parallel stability analysis and N-periodic SPOD, it is possible to provide a comprehensive characterisation of the linear mechanisms that arise in hypersonic streaky boundary layers and how these are impacted by streak amplitude.
}

The paper is organised as follows. Sec.~\ref{sec:sim-fmwk} introduces the simulation framework with a description of the flow conditions and the numerical methods. In Sec.~\ref{sec:lst-tg} a spatial optimal disturbance is obtained and streaky baseflows are computed as the non-linear response to this disturbance. Next, in Sec.~\ref{sec:2d-lst} a local stability analysis is performed in the spatial framework on the 2D baseflow and the various linear growth mechanisms are characterised. Having discussed the stability properties of the streaky baseflows a set of linearised DNS are performed in Sec.~\ref{sec:ldns} to evaluate the response of a streaky baseflow to white noise for varying streak amplitudes. The datasets are decomposed using the N-periodic SPOD and the modal decomposition is compared with the results of the local analysis in order to characterise the structures observed in the simulations. Finally, in Sec.~\ref{sec:conclusion} the streak effects are discussed and perspective perspectives for future research are proposed.

\section{Simulation Framework}
\label{sec:sim-fmwk}

\subsection{Case description}
\label{sec:b-l}

\begin{figure}
    \centering
    \includegraphics[width=\textwidth]{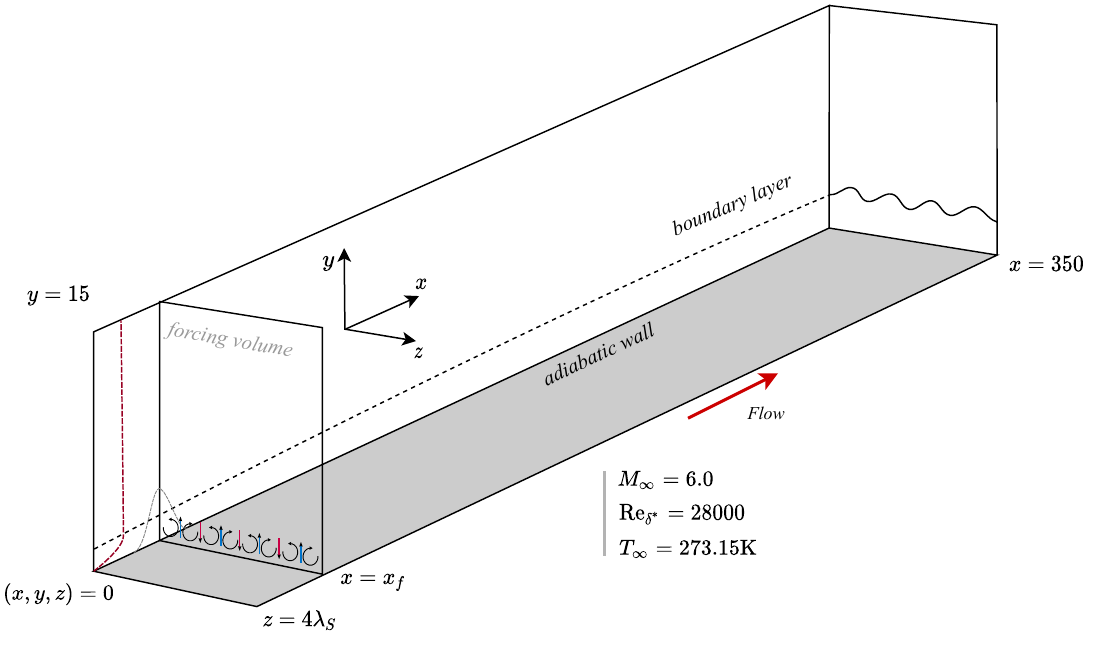}
    \caption{General view of the simulation domain considered. Dotted line : boundary-layer height~$\delta_{99}$}
    \label{fig:dns_domain}
\end{figure}
To define the flow conditions, we follow the successive studies of \citet{detullioInfluenceBoundarylayerDisturbances2015}, \citet{vandeneyndeNumericalSimulationsTransition2015} and \citet{lefieuxDNSStudyRoughnessInduced2019}. A hypersonic viscous flow at $M_\infty=6.0$ and $\text{Re}_{\delta^*} = 28 000$ over a flat plate is considered. The freestream temperature is set at $T_\infty=273.15$K and the plate wall is adiabatic. A compressible self-similar boundary-layer profile is directly imposed at the domain inlet at $x=0$. The inlet displacement thickness $\delta^*$ at $x=0$ is used as the reference length $L_{\text{ref}}$. At the inflow, the boundary layer thickness is $\delta_{99} = 1.35\delta^*$.

With these conditions, the computational domain is designed to contain four streaks of spanwise wavelength $\lambda_S$. The wavelength value will be further defined in Sec.~\ref{sec:lst-tg}. This number of streaks is chosen after following the study of \citet{anderssonBreakdownBoundaryLayer2001} \& \citet{paredesInstabilityWaveStreak2019} suggesting a prevalence of subharmonic instabilities. The domain has a dimensions, $L_x\times L_y \times L_z = 350 \times 15 \times 4\lambda_S $. Three sponge zones  at the upper boundary, inlet and outlet, are added to ensure clean outflow conditions for which additional details are given in section~\ref{sec:num-bc}. A general view of the geometry is given in Fig.~\ref{fig:dns_domain}. The domain is discretised in a structured fashion with $N_{\text{pts}} = N_x \times N_y \times N_z = 4000 \times 300 \times 300$ points. The number of points and the distributions are chosen considering the mesh convergence of \citet{lefieuxDNSStudyRoughnessInduced2019} for identical flow conditions and similar numerical methods.

\begin{table}[h]
    \centering
    \begin{tabular}{c c c c c c }
        $M_\infty$ & $Re_{\delta^*}$ & $T_\infty$ (K) & Pr & $T_{suth}$ (K) & $C_p$ $\text{J} \text{kg}^{-1} \text{K}^{-1}$\\ \hline \hline
         $6.0$ & $28~000$ & $273.15$ & $0.72$ & $273.15$ & $1004.0$ \\ \hline
    \end{tabular}
    \caption{Summary of the considered flow conditions}
    \label{tab:my_label}
\end{table}



\subsection{Governing equations}

The compressible Navier-Stokes equations are solved on the considered domain for an initial condition $\pmb q_0$ and a set of boundary conditions. The vector $\pmb q = (\rho , \rho \pmb u , \rho E)^T$ being the conservative state vector and $\pmb u = (u_x,u_y,u_z)^T$ the cartesian velocity vector. The variables $\rho$, $E$ are respectively the density and the total energy per unit mass. For computation purposes, the non-linear system is solved in a generalised-coordinate framework. Thus, we consider the usual set of cartesian coordinates $\bm{\mathsf{X}} = (x,y,z)$ and a curvilinear set of coordinates defined on the structured mesh $\mathcal{X} = (\xi, \eta, \zeta)$, conservative transformation metrics are computed as in \citet{dengFurtherStudiesGeometric2013}. Using a compact partial derivative notation $\partial . / \partial \xi = \partial_\xi$ the system equations read,
%
\begin{align}
        \partial_t \hat{\pmb q} = \partial_\xi(\pmb F_\xi - \pmb F_\xi^v)  \label{eq:NS} 
                            + \partial_\eta(\pmb F_\eta - \pmb F_\eta^v)
                            + \partial_\zeta(\pmb F_\zeta - \pmb F_\zeta^v). 
\end{align}
Where $\hat{\pmb q} = {\pmb q}/{J}$ is the conservative state vector locally scaled by the metric Jacobian ${J}=\det(\bm{\mathsf J}_{\mathsf{X} \rightarrow \mathcal{X}})$. The flux-vectors $\pmb F_m$ and $\pmb F_m^v$, $m= \xi,\eta,\zeta$, respectively define the inviscid and viscous fluxes of the Navier-Stokes equations. The former fluxes are defined as,
\begin{align}
    \pmb F_m = \frac{1}{J}
    \begin{pmatrix}
    \rho \pmb v^m \\
    \rho \pmb u_n \pmb v^m + p \nabla_x^m \\
    (\rho e + p) \pmb v^m
    \end{pmatrix}.
\end{align}
The cartesian velocity vector $\pmb u$ is used to define the contravariant velocity vector $\pmb v$ in $\mathcal{X}$ space, with $n=x,y,z$,
\begin{align}
    v^\xi = \pmb u \cdot \nabla_{\sf x} \xi, && 
    v^\eta = \pmb u \cdot \nabla_{\sf x} \eta, &&
    v^\zeta = \pmb u \cdot \nabla_{\sf x} \zeta.
\end{align}
With the transformation metrics for a given direction $(\bullet)$ defined as $\nabla_{\sf{x}}\bullet = (\partial_x \bullet, \partial_y \bullet, \partial_z \bullet)^T$. 
Viscous-fluxes $F_m^v$ are expressed as,
\begin{align}
    \pmb F_m^v = \frac{1}{J}
    \begin{pmatrix}
    0 \\
    \mu\underline{\pmb{\tau}} . \nabla_{\sf x}^m \\
    \pmb u . \underline{\pmb{\tau}} . \nabla_{\sf x}^m - \pmb \phi . \nabla_{\sf x}^m
    \end{pmatrix}.
\end{align}
And we define the viscous stress tensor in $\mathcal{X}$ space as the contraction $\underline{\pmb{\tau}} . \nabla_{\sf x}^m $. The viscous stress tensor $\underline{\pmb \tau}$ and the heat flux vector $\pmb \phi$ are defined in $\bm{\mathsf{X}}$ space as follows,

\begin{align}
    \underline{\pmb \tau} &= \left( \nabla_{\sf x} \pmb u + \nabla_{\sf x} \pmb u^T - \frac{2}{3}(\nabla_{\sf x} \cdot \pmb u) \pmb I \right), \\
    \pmb \phi &= \kappa \nabla_{\sf x} T .
\end{align}

The physical quantities used are $\mu(T)$, the dynamic viscosity linked to the static temperature $T$ through the Sutherland law and $\kappa$, the thermal conductivity defined by the Fourier law. The relatively low enthalpy levels considered allow us to use the ideal gas equation of state: $p = \rho r_{gas} T$, relating the pressure $p$ to the temperature $T$ and density $\rho$ with the perfect gas constant $r_{gas}$.


\subsection{Numerical methods} \label{sec:num-bc}
The computations are performed with our in-house Direct Numerical Simulation (DNS) code \texttt{CurviCreams}, dedicated to the study of hypersonic flows in generalised coordinates. The Navier-Stokes equations are solved as expressed in Eq.~\ref{eq:NS} in a finite-difference framework. As Hypersonic flows contain shocks, the numerical methods must distinguish smooth and non-smooth regions of the flow. For smooth regions, the flux vectors $\pmb F_m$ and $\pmb F_m^v$ are discretised using a centered, 9-point, explicit scheme of order 8. 
Due to their low dissipative nature, high-order schemes are prone to numerical oscillations. This issue is addressed by the addition of a tailored dissipation 
through explicit 9-point optimised filters, as defined by \citet{bogeyFamilyLowDispersive2004}. The scheme used effectively resolves the wavenumber content down to five points per wavelength. At the borders, the stencils progressively switch to a de-centred second-order scheme at the boundary point. This order reduction is chosen to increase the computation stability at the boundaries for the high gradients considered. The solution is time marched explicitly with a third order, low-storage and Total-Variation-Diminishing, Runge-Kutta (RK3-TVD) as defined by \citet{gottliebTotalVariationDiminishing1998}

For non-smooth regions of the flow and strong discontinuities, a shock-capturing scheme suited for the generalised coordinates framework is used. The shock-capturing dissipation $\mathcal{D}_s$ is based on the Adaptive Non-Linear Artificial Dissipation (ANAD) developed by \citet{kimAdaptiveNonlinearArtificial2001} and also similar to \citet{sciacovelliAssessmentHighorderShockcapturing2021}. The method consists in a conservative formulation of a localised dissipation term, computed through a sensor using high-order filters, evaluating the flow smoothness on the pressure field. The dissipation term obtained is multiplied by a local shock sensor $\Theta_s$ using the formulation of \citet{bhagatwalaModifiedArtificialViscosity2009} to avoid unnecessary smoothing of transitional structures in the boundary layer. 





\subsection{Discrete Linearisation}
A discrete linearisation procedure is implemented to compute the full Jacobian or to perform Linear DNS (LDNS) by time marching of the linearised operator. The computation of the Jacobian of the non-linear operator is done in the spirit of the works of \citet{fosasdepandoEfficientEvaluationDirect2012} and \citet{beneddineCharacterizationUnsteadyFlow2017} and is described below in our context. 

Denoting the non-linear discrete Navier-Stokes operator of Eq.~\ref{eq:NS} by $\bm{\mathsf N}(\pmb q)$ and $N_{dof}$ the number of degrees of freedom, the time  evolution of any state vector $\pmb q(x,t) \in \mathbb{R}^{N_{dof}}$ computed by the DNS solver then reads:

\begin{align}
    \frac{\partial \pmb q}{\partial t} = \bm{\mathsf N}(\pmb q). \label{eq:nolin-ns}
\end{align}

The LDNS is performed by first assuming a decomposition of the state vector $\pmb q(x,y,z,t)$ in a steady base state $\bar {\pmb q}(x,y,z)$ such that $\bm{\mathsf N}(\bar{\pmb q}) = 0$ and an unsteady disturbance field of small amplitude $\pmb q'(x,y,z,t)$ such that for any point $(x,y,z)$ in the domain : $\pmb q (t) = \bar {\pmb q} + \pmb q'(t)$. The operator $\bm{\mathsf N}$ is then linearised around the base state $\bar{\pmb q}$ and the linear system Jacobian $\bm{\mathsf L}_{\bar{\pmb q}}$ reads,

\begin{align}
    \bm{\mathsf L}_{\bar{\pmb q}} = \left.\frac{\partial\bm{\mathsf N}(\pmb q)}{\partial \pmb q}\right|_{\bar{\pmb q}}. 
\end{align}

Considering that the DNS code can compute $\bm{\mathsf N}(\pmb q)$ for any given $\pmb q$, the action of its tangent operator $\bm{\mathsf L}_{\bar{\pmb q}}$ linearised around $\bar{\pmb{q}}(x,y,z)$ can then be extracted through a first-order approximation,
\begin{align}
    \label{eq:1s-order-L}
    \bm{\mathsf L}_{\bar{\pmb q}}\pmb q' = \frac{\bm{\mathsf N}(\bar{\pmb q} + \varepsilon \pmb q') - \bm{\mathsf N}(\bar{\pmb q})}{\varepsilon},
\end{align}
where the linearised dynamics of a disturbance vector $\pmb q'$ around $\bar{\pmb q}$ are obtained integrating,

\begin{align}
    \frac{\partial {\pmb q}'}{\partial t} = \bm{\mathsf L}_{\bar{\pmb q}}\pmb q'. \label{eq:lin_dyn}
\end{align}

The linearised problem is time marched using the same schemes and boundary conditions as the non-linear problem, with the same computational cost, allowing rigorous comparisons between LDNS and DNS results. This first-order linearisation is known to be sensitive to the small step $\varepsilon$. Following previous studies using this numerical procedure with double-precision arithmetic \citep{bugeat3DGlobalOptimal2019, browneSensitivityAnalysisUnsteady2014}, a value of $\varepsilon=10^{-8}$ has been selected.

\section{Building streaky baseflows} \label{sec:lst-tg}

A canonical flat-plate baseflow is obtained by time marching the boundary layer in the domain until $\bm{\mathsf N}(\pmb q) \leq 10^{-8}$. From this initial baseflow, to obtain the streaky baseflow, an exogenous body force is added close to the boundary. This forcing is determined based on a locally parallel spatial transient growth analysis. The choice of producing the streak by the introduction of a vortical optimal disturbance is motivated by the idea of having a controlled and reproducible streak amplification within the domain as optimal disturbances provide an upper bound of streak growth. The two following sections give the details of the procedure.

\subsection{Homogenous boundary layer baseflow}
With the numerical tools and freestream flow conditions defined above, a laminar 2D boundary layer is computed on the adiabatic flat plate of Fig.~\ref{fig:dns_domain} with the DNS code. The simulation is initiated with the self-similar boundary-layer profile imposed in the whole domain and the computation is time marched for around 80000 iterations until convergence with $\bm{\mathsf N}(\pmb q)<10^{-8}$. 

\begin{figure}
    \centering
    \includegraphics[height=7cm]{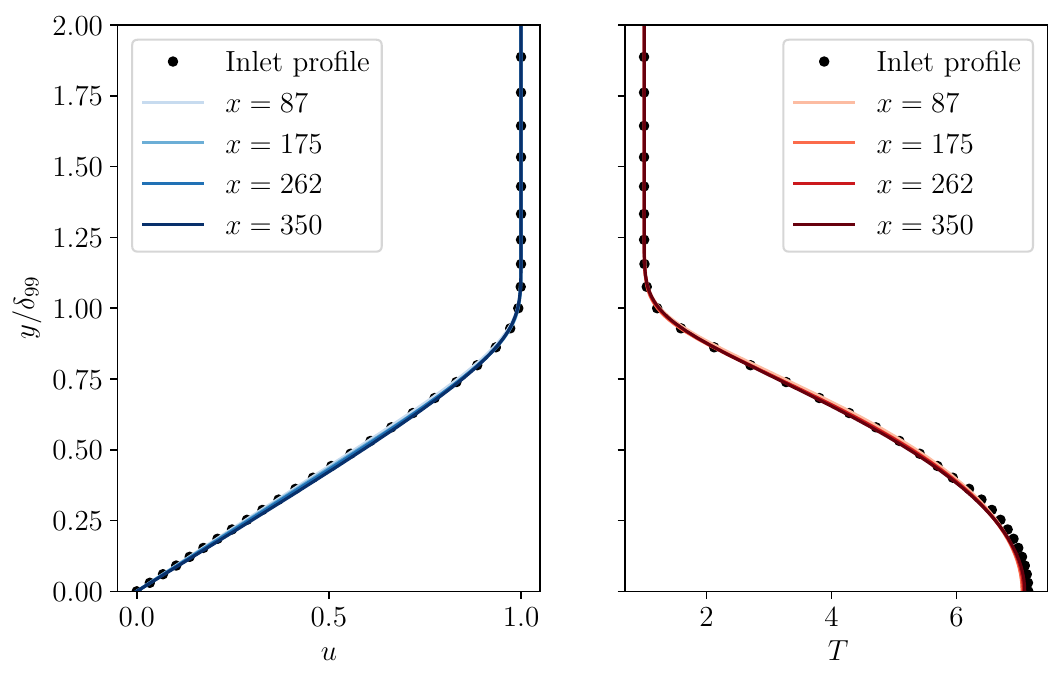}
    \caption{Self-similarity of the initial laminar boundary layer over the adiabatic flat-plate }
    \label{fig:self_sim}
\end{figure}

This first computation leads to a fully developed laminar boundary layer over an adiabatic wall. The baseflow self-similarity can be verified in Fig.~\ref{fig:self_sim} for the axial velocity and temperature. A boundary layer profile in conservative variables is extracted at $x=5$ to build the local stability problem. 

\subsection{Optimal disturbance computation}



The 1D LST code computes the optimal disturbance $\pmb f_{S}$ and its response $\pmb q'_S$. It uses the compressible linearised Navier-Stokes equations, such as written in \citet{mackBoundaryLayerLinearStability1984} to build eigenvalue problems. The method is based on a spectral collocation discretisation through Chebyshev polynomials. With this linear operator, steady streaks disturbances are sought for an optimal spanwise-wavenumber $\beta = 2\pi / \lambda_z$. Thus, eigenvalue problems are solved for increasing values of $\beta\in[0,0.4]$ and $\omega=0$. For each $\beta$ value, a spatial transient growth analysis is performed to find the disturbance and response maximizing the spatial gain $G(x)$. Maximum gain is searched for $x=350$ such that streaks are continuously growing and reach maximum amplitude exactly at the end of the domain with the following gain definition.  
\begin{align}
    G = \frac{||\pmb q'_S(x=350)||_E}{||\pmb f_S(x=5)||_E},
\end{align}
with the compressible energy norm $||.||_E$ defined by \citet{chuEnergyTransferSmall1965} for which more details are given in Sec.~\ref{sec:floq-spod}. The procedure used to compute the optimal disturbances from the spatial eigenvalues spectrum follows the work of \citet{jordanModalNonmodalLinear2017} where upstream-travelling waves are filtered out using the Briggs-Ber criterion \citep{briggsElectronStreamInteractionPlasmas1964}.

\begin{figure}
    \begin{subfigure}[t]{0.45\textwidth}
        \centering
        \includegraphics[height=0.3\textheight]{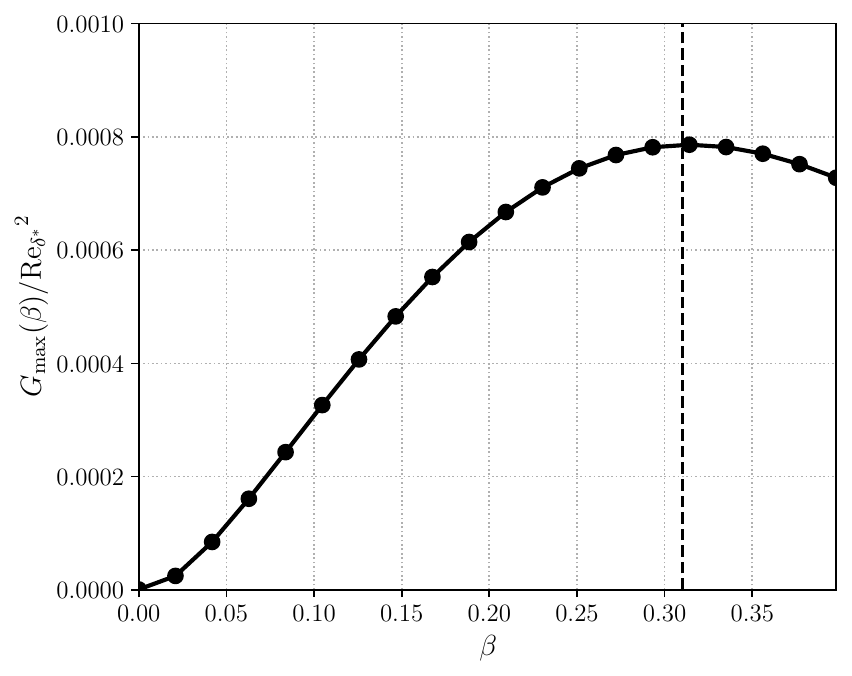}   
        \caption{Evolution of the optimal gain G for various values of $\beta$}  
        \label{fig:gains-qf1}   
    \end{subfigure}
    \hfill
    \begin{subfigure}[t]{0.45\textwidth}
        \centering
        \includegraphics[height=0.3\textheight]{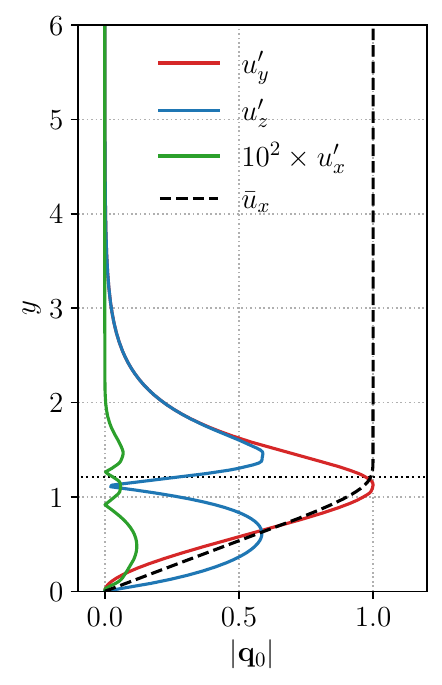} 
        \caption{Eigenfunctions of the optimal forcing vector $\hat{\pmb f}_S$. Dashed : baseflow streamwise velocity profile}
        \label{fig:gains-qf2}
    \end{subfigure}
    \caption{Transient growth gain evolution and associated optimal disturbance}
    \label{fig:gains-qf}
\end{figure}

The calculated gain evolution as a function of $\beta$ is given in Fig.~\ref{fig:gains-qf}. An optimal spanwise-wavenumber is found at $\beta_{opt}=0.31$ and leads to a gain scaled by $\text{Re}_{\delta^*}$ of $G=0.008$. These values are in good agreement with the transient-growth values obtain with PSE by \citet{paredesOptimalGrowthHypersonic2016} at the same Mach number. The linear steady disturbance and response can therefore be written as, 
\begin{align}
    \pmb f_S'(y,z) = \hat{\pmb f}_S(y) e^{i \beta_{opt} z}, && \pmb q_S'(y,z) = \hat{\pmb q}_S(y) e^{i \beta_{opt} z}.
\end{align}

The velocity profiles of the optimal vortical disturbance $\hat{\pmb f}_{S}(y)$ associated with the maximum gain at $\beta_{opt}=0.31$ are depicted in Fig.~\ref{fig:gains-qf2}. The $u_y$ and $u_z$ components can be seen and the core of the associated streamwise vortex is located just below the boundary layer edge, in agreement with previous studies. The axial velocity component $u_x$ is also shown and is two orders of magnitude smaller than $u_y$ and $u_z$, showing the optimal disturbance to comprise a roll which will lead to large amplification via the lift-up mechanism. 

\subsection{Steady streak forcing}

A steady forcing is imposed to generate the streaks. Using the forcing vector $\pmb f_S (\pmb x,t)$ of initial amplitude $A_0$, the homogenous boundary layer state vector is disturbed in a control volume defined by a spatial distribution function $\phi_f(\pmb x)$. This function is of unit integral in the $x$-direction and is defined on a number of points in the streamwise direction sufficient for the 9-point spatial scheme to resolve the deposed energy. Adding a forcing $\pmb f_S$ to the non-linear system of Eq.~\ref{eq:nolin-ns} reads,
\begin{align}
    \frac{\partial \pmb q}{\partial t} = \bm{\mathsf N}(\pmb q) + \pmb f_S , \label{eq:forced-nolin-ns}
\end{align}
with the vector $\pmb f_S$ expressing the disturbance taking the form,
\begin{align}
    \pmb{f}_S(\pmb x,t) =  A_0 \phi_f (x) \hat{\pmb f}_S(y) e^{i\beta_{opt} z}. \label{eq:disturb}
\end{align}
The smooth distribution of Eq.~\ref{eq:space_distrib} is tailored to locate most of $\pmb f_S$ energy in a section of the domain restricted between $x_0=4$ and $x_1=6$, as depicted in Fig.~\ref{fig:dns_domain}, with a dotted line,
%
\begin{align}
    \phi_f(x) = \frac{1}{2}\frac{(\tanh(\sigma(x_0 - x)) - \tanh(\sigma(x_1 - x)))}{|x_1 - x_0|}. \label{eq:space_distrib}
\end{align}
%



The base states are obtained from time marching at a constant CFL number rather than a constant time-step $\Delta t$. Furthermore, the amplitude of the stationary forcing can be substantial when seeking high-amplitude streaks. Thus, a linear amplitude ramp for $A_0$ can be used to avoid strong discontinuities during the transient stage of the simulation when propagating the front of the forcing downstream.

\subsection{Baseflows properties}

\begin{figure}
    \centering
    \includegraphics[width=\textwidth]{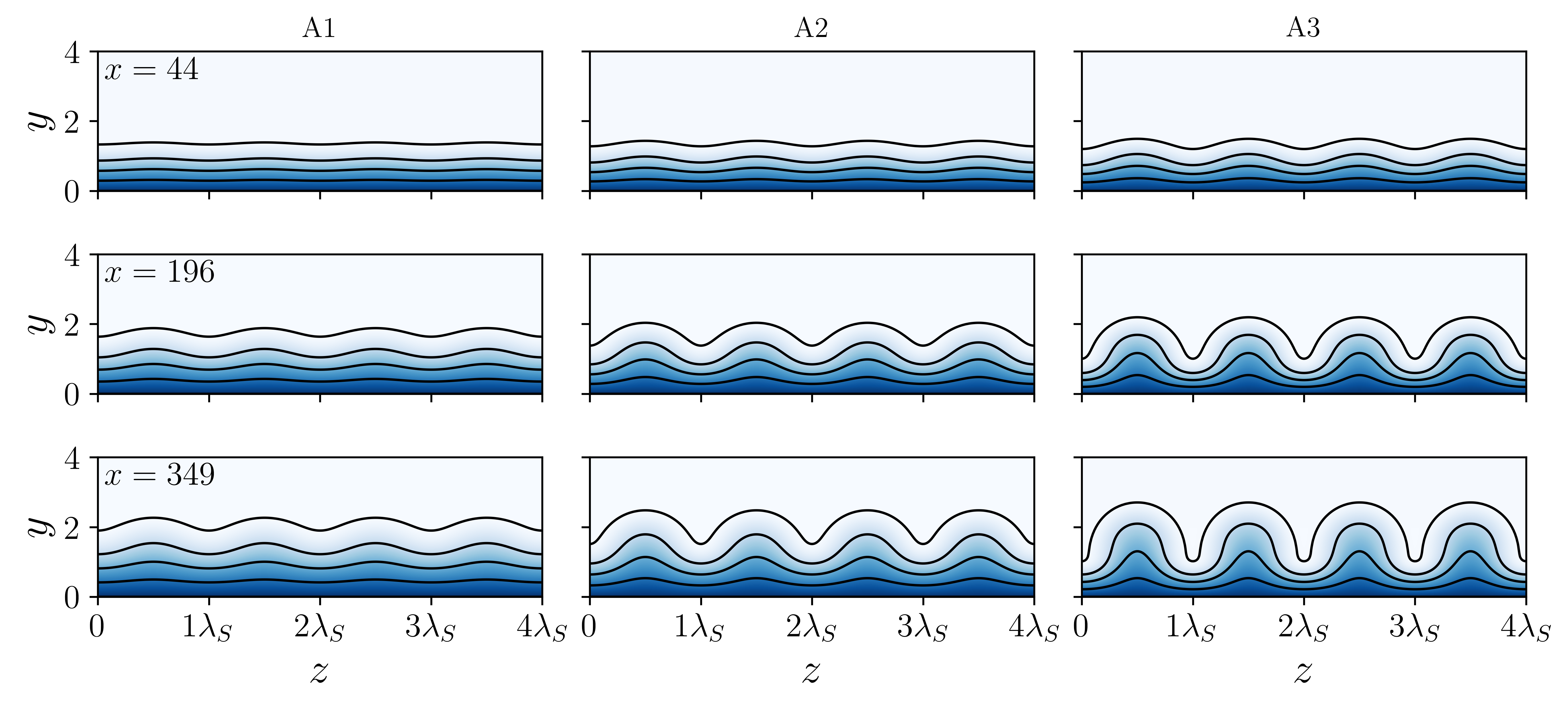}
    \caption{Baseflows evolution for A1 (left), A2 (middle) and A3 (right). Normalised contours of $u_{x}$ streamwise velocity, five iso-contours in $[0,1]$.}
    \label{fig:stk-bf}
\end{figure}

Using the forcing procedure, three separate computations are carried out with increasing values of initial amplitude $A_0$. These three cases will be referred to as A1, A2 and A3, corresponding respectively to low- medium- and high-amplitude streaks in the homogeneous boundary layer. The baseflow associated with these cases is denoted $\bar{\pmb q}_S$. Whereas the baseflow associated with the undisturbed homogeneous boundary layer is denoted $\bar{\pmb q}_U$. The non-linear steady streaks can be retrieved with $\pmb q'_S = \bar{\pmb{q}}_S - \bar{\pmb q}_U$. The associated initial amplitude and final streak amplitude are given in Tab.~\ref{tab:stk_asu}. To compute these streak amplitude $A_{s}$, we follow the definition of \citet{anderssonBreakdownBoundaryLayer2001} using the streak axial velocity $u'_{x,S}$. 
\begin{table}
    \begin{center}
            \begin{tabular}{l c c c }
                     & A1 & A2 & A3 \\
                     & \textit{low} & \textit{medium} & \textit{high} \\  \hline
               $A_0$ & $5.6\times10^{-3}$ & $1.5\times10^{-2}$ & $2.8\times10^{-2}$   \\
               $A_{s}(x=350)$ & $0.082$ & $0.217$ & $0.383$ \\ \hline
            \end{tabular}
        \caption{Summary of the forcing initial amplitudes and the associated response amplitudes}
        \label{tab:stk_asu}
    \end{center}    
\end{table}

\begin{align}
    A_{s}(x) = \frac{1}{2} \left[ \max_{y,z}\left(\pmb u_{x,S}'(x)\right) - \min_{y,z}\left(\pmb u_{x,S}'(x)\right)\right]. \label{eq:Asu}
\end{align}
For the cases A1, A2 \& A3, the two-dimensional axial velocity profiles obtained are depicted in Fig.~\ref{fig:stk-bf}. In each corresponding column, the streak amplitude growth is shown at three streamwise positions of the domain $x\in[44,196,349]$. The progressive boundary-layer distortion induced by the lift-up effect can be clearly seen with the low-speed streaks moving up and the high-speed streaks moving closer to the wall.

\begin{figure}
    \centering 
    \begin{subfigure}[t]{0.46\textwidth}
        \includegraphics[width=\textwidth]{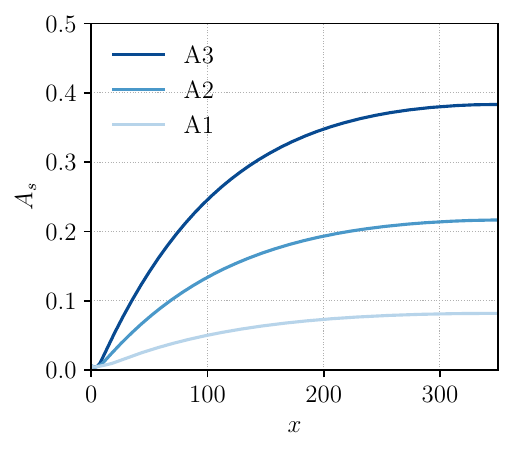}
        \caption{Evolution of $A_s(x)$ for the three considered initial amplitudes}
        \label{fig:streaks_asu}
    \end{subfigure}
    \hfill
    \begin{subfigure}[t]{0.49
        \textwidth}
        \includegraphics[width=\textwidth]{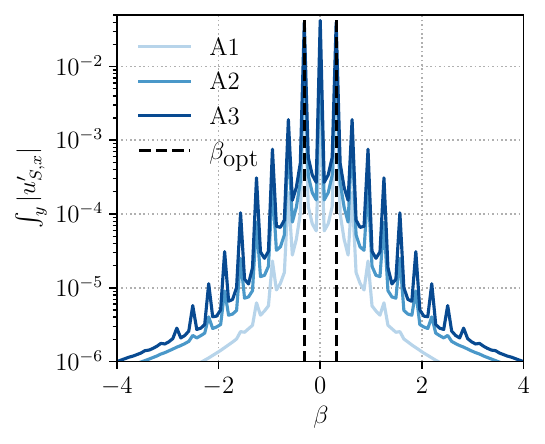}
        \caption{Comparison of the integrated spanwise spectra of $u_{S,x}'$ at $x=350$. Dashed line is  $\beta_{\text{opt}}$ of the optimal linear response}
        \label{fig:stk_bf_fft}
    \end{subfigure}
    \\
    \vfill
    \begin{subfigure}[t]{\textwidth}
        \includegraphics[width=\textwidth]{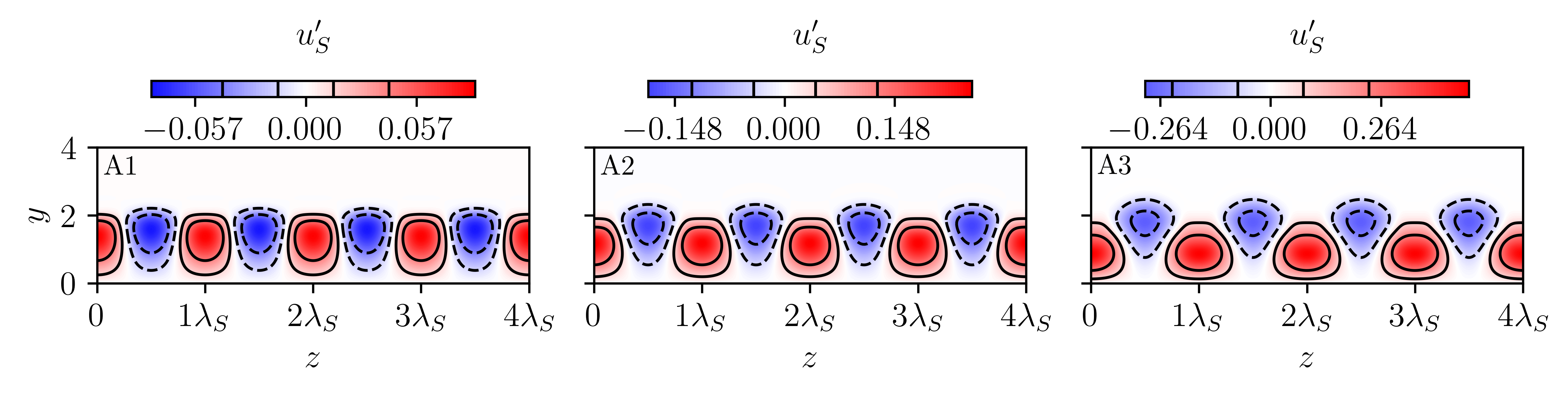}
        \caption{Contours of $\pmb{u}'_{S}$ at $x=350$. From left to right : $A1$, $A2$, $A3$}
        \label{fig:streaks_shape}
    \end{subfigure}
    \caption{Streaks amplitude evolution and shape modification with increased forcing strength}
    \label{fig:streaks_x}
\end{figure}

Associated with these baseflow profiles, the streak-amplitude evolution is shown in Fig.~\ref{fig:streaks_x} along with the final streak shape at $x=350$. For each case, the amplitude growth in Fig.~\ref{fig:streaks_asu} is monotonic and reaches a maximum close to the end of the domain as expected from the optimal disturbance computation. The associated non-linear responses from the linear optimal disturbance are shown in Fig.~\ref{fig:streaks_shape} with contours of $\pmb{u}'_{x,S}$. The observed streak shapes reveal a progressive distortion of the high and low-speed disturbances as the forcing amplitude increases. This distortion is associated with increased non-linear effects and saturation of the initial linear streak. The increased number of energetic spanwise harmonics indicates that saturation excites these additional spanwise harmonics through non-linearity. This energy redistribution among harmonics of $\beta_{opt}$ results in a shape change \citep{ranModelingModeInteractions2019}. The saturation effect is illustrated in Fig.~\ref{fig:stk_bf_fft} with a spanwise Fourier transform of the velocity disturbances. In this figure, the increased number of $\beta_{opt}$ harmonics between A1 and A3 can be noticed. Additionally, the streak amplitude saturation is also observable with the respective amplitudes differences between A1, A2 \& A3. 

Having obtained these baseflows, the effects of streak distortion and their growth is investigated in the next section. We first explore the 2D-local stability characteristics of the streaky baseflows before investigating the linear response to white-noise forcing.


\section{Spatial 2D local stability analysis} \label{sec:2d-lst}


Although the streak baseflow is 3D in nature, it presents a slow development in the streamwise direction. Therefore, a choice is made to use a locally parallel approximation in order to perform 2D spatial analyses of the boundary layer. This approach proved to be efficient in correctly estimating convective instabilities properties in similar flows \citep{schmidtViscidInviscidPseudoresonance2014, monteroAnalysisInstabilitiesInduced2021, detullioInfluenceBoundarylayerDisturbances2015}, and it has the added virtue of providing a convenient classification of the different instability mechanisms based on the local eigen spectrum.

The local stability analysis is intended to support the global LDNS database by identifying and naming the instability mechanisms supported by the streaky boundary layer base flow. The focus is mainly set on the first-mode waves at low-frequency as this instabilitly mechanism received only little attention for streaky hypersonic boundary layers. First, the effect of a very small baseflow distortion is studied. Subsequently, the trajectories of the unstable eigenvalues are tracked in the complex wavenumber plane as a function of streamwise position. 

\subsection{Notation}

In what follows, the numerous instabilities that arise are studied. To facilitate clear discussion, a notation system is introduced to distinguish the different instabilities. In hypersonic boundary layers, the two main linear, modal growth mechanism are known to be the first- and second-mode convective instabilities \citep{mackBoundaryLayerLinearStability1984}. In the local framework, such instabilities are related to the so-called slow and fast eigenmodes of the linear operator, referred as S and F respectively \citep{fedorovPrehistoryInstabilityHypersonic2001}. With modulation of the baseflow by streaks, these mechanisms are modified by the streaky base flow; and, furthermore, a third family of convectively unstable waves is found to arise which we refer to as streak-dependent instabilities. We denote these three families :
\begin{itemize}
    \item FM : the family of first-mode instabilities ;
    \item MM : the family of second-mode instabilities, as for Mack-modes ;
    \item SM : the family of streak-dependant instabilities. 
\end{itemize} 
\newcommand{\FM}[2]{\text{FM}^{#1}_{#2}}
\newcommand{\MM}[2]{\text{MM}^{#1}_{#2}}
\newcommand{\SM}[2]{\text{SM}^{#1}_{#2}}
In addition to these families, it has been demonstrated that streaks impart a specific geometrical organisation to the instabilities that they support \citep{anderssonBreakdownBoundaryLayer2001}. These properties are manifest in the symmetric/antisymmetric nature of the eigenfunctions with respect to the streak structure. Along with these symmetries, the instability spanwise wavelength, $\lambda_z$ can extend over one, two or more streaks. Therefore, the value of $\lambda_z$ with respect to the streak spanwise wavelength $\lambda_S$ is also a key parameter and is added in the notation.

Instabilities are thus denoted $(\bullet)_{s/a}^k$, where $s$ and $a$ respectively indicate the symmetric and anti-symmetric organisation of the eigenfunction around the low-speed streak centerline, and, $k$ denotes the spanwise wavelength relation between the instability eigenfunctions and the streaks such that $\lambda_z = k \lambda_S$. To follow the terminology of previous studies \citep{paredesTransitionDueStreamwise2016}, this wavelength relationship will be referred to as fundamental for $k=1$ or subharmonic for $k=2$. Finally, as an example, $\text{MM}^2_s$ would denote an symmetric, subharmonic $(\lambda_z=2\lambda_S)$ second-mode instability. 
\subsection{Local stability framework}


The baseflows obtained with the streaks amplitudes A1, A2 \& A3 are fixed points of the Navier-Stokes equations. Therefore, a spatial stability problem is built with the linear-disturbance equation of Eq.~\ref{eq:lin_dyn} around the three base states. Using a locally parallel hypothesis, the disturbance ansatz is,
\begin{align}
    \pmb q'(y,z,t) = \hat{\pmb q}(y,z) \exp (i\alpha x - \omega t), 
\end{align}
with $\hat{\pmb q}(y,z)$ the 2D eigenfunctions, $\omega=2\pi f$ the circular frequency and $\alpha$ the axial wavenumber. Using this ansatz in Eq.~\ref{eq:lin_dyn}, after some developments, we obtain the local stability problem refactored in matrix form \citep{mackBoundaryLayerLinearStability1984},
\begin{align}
    \omega \bm{\mathsf L_0}\pmb q’ = \left( \bm{\mathsf F_0} + \alpha \bm{\mathsf F_1} + \alpha^2 \bm{\mathsf F_2} \right)\pmb q’ . \label{eq:stk_spatial_stability_pb}
\end{align}
The spatial stability problem is obtained by imposing a given value of $\omega\in\mathbb{R}$ and solving the polynomial eigenvalue problem~\ref{eq:stk_spatial_stability_pb} for  $\alpha\in\mathbb{C}$. This formulation gives the spatial growth rate $\Im(\alpha)$ and the associated eigenfunctions $\hat{\pmb q}$ of disturbances evolving around baseflows slices $\pmb q_S(y,z)$. The different linear operators are obtained using the code developed by \citet{schmidtViscidInviscidPseudoresonance2014}. The use of 4th-order explicit finite-difference schemes leads to sparse operators. From these operators, the quadratic eigenvalue problem is cast in an augmented eigenvalue problem to obtain an equivalent linear formulation such that,
\begin{align}
    \underbrace{\begin{bmatrix}
        \omega\bm{\mathsf L_0} - \bm{\mathsf F_0} & \bm{\mathsf 0} \\ 
        \bm{\mathsf 0} & \bm{\mathsf I}
    \end{bmatrix}}_{\bm{\mathsf L}}
    \begin{bmatrix}
        \pmb q' \\ \alpha \pmb q'
    \end{bmatrix}
    =
    \alpha
    \underbrace{\begin{bmatrix}
        \bm{\mathsf F_1} & \bm{\mathsf F_2}\\ 
        \bm{\mathsf I}  & \bm{\mathsf 0}
    \end{bmatrix}}_{\bm{\mathsf M}}
    \begin{bmatrix}
        \pmb q' \\ \alpha \pmb q'
    \end{bmatrix}. \label{eq:2d_lst_GEVP}
\end{align}
The problem is solved on a distributed architecture with an iterative Krylov-Shur method and an LU decomposition through the SLEPc and MUMPS libraries. A shift-invert strategy is used to target the eigenvalue search to a specific location $\tau$ of the complex plane. The target $\tau$ is located in the region defined by planar unstable disturbances with $\Im(\alpha)<0$, having phase speed $c_r \in [1-1/M_\infty, 1+1/M_\infty]$. This specific region of the complex plane is known to contain the discrete eigenvalues S and F, associated with the first-mode and second Mack-mode instabilities \citep{fedorovPrehistoryInstabilityHypersonic2001}. 

The following computations are made in a domain of size $y\times z \in [0,40]\times[0,2\lambda_s]$ discretised by $N_y\times N_z = 200 \times 150$ points. The stretching of \citet{hanifiTransientGrowthCompressible1996} is used in the $y$ direction with the stretching factor $y_i=1.5$. Dirichlet boundary conditions imposing $\pmb q'=0$ are used at $y=0$ and $y=y_{\max}$ and the $z$ direction is considered periodic.

\begin{figure}[t]
    \centering
    \centerline{\includegraphics[width=\textwidth]{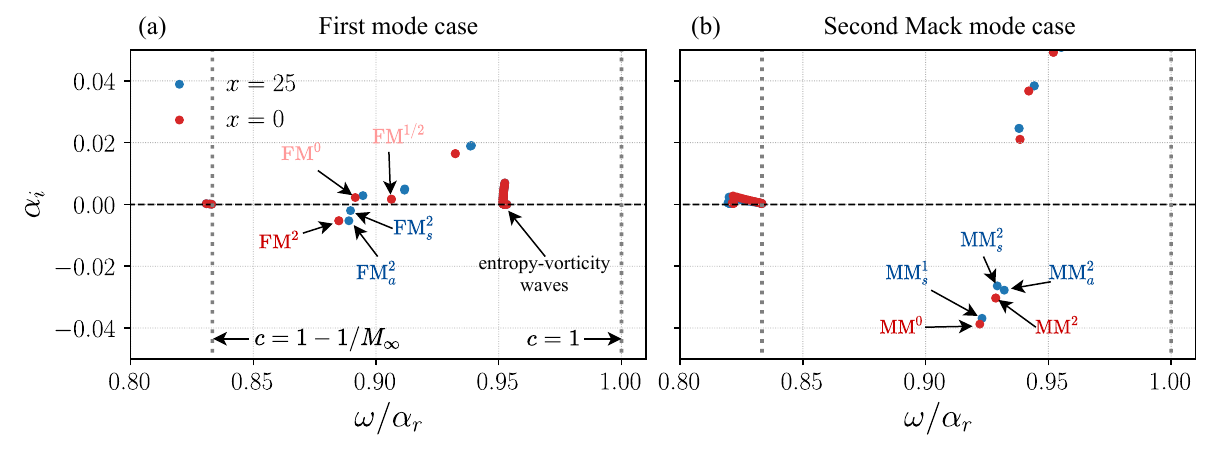}}
    \caption{Comparison of spatial spectra at frequencies $f=0.1$ (left) \& $f=0.32$ (right) between $x=0$ (red) and $x=25$ (blue) for A1. The $N_\alpha=50$ closest eigenvalues from $\tau$ are drawn}
    \label{fig:2d_lst_x1_x20}
\end{figure}
%
\subsection{Effect of small distortions}
At first, the effect of a small $(\approx1\%)$ boundary layer distortion is investigated to understand the effect of streaks on the first- and second-mode instabilities. Spectra at two streamwise stations are computed and shown in Fig.~\ref{fig:2d_lst_x1_x20} in the phase-speed and growth rate frame $(c_{\phi}, \alpha_i)$ with $c_{\phi}=\omega/\alpha_r$. Red eigenvalues relate to the undisturbed boundary layer at $x=0$. Blue eigenvalues correspond to the streaky baseflow A1 at $x=25$. At this station, A1 has a streak amplitude $A_{s}=0.016$. For each station, spectra are computed for the first mode frequency $f=0.1$ (left) and the second-mode frequency $f=0.32$ (right).


Unstable eigenvalues are found in the interval $c_\phi\in[1-1/M_\infty,1]$ for $\alpha_i<0$. Recalling the aforementioned notation, FM for first-mode instabilities and MM for second-mode instabilities, we observe for the frequency $f=0.1$ in Fig.~\ref{fig:2d_lst_x1_x20}a, one unstable mode FM at $x=0$. This mode is identified through its eigenfunction and is oblique with $\lambda=2\lambda_S$. Additional stable discrete eigenvalues are found for eigenmodes of wavelength $\lambda\in[0,1/2\lambda_S]$. These discrete eigenvalues correspond to the first mode branch with different spanwise-wavenumbers, $\lambda_z=k\lambda_S, \  k\in[0,1,2,...]$, allowed by the finite and periodic domain. 

\begin{figure}[t]
    \centering
    \centerline{\includegraphics[width=\textwidth]{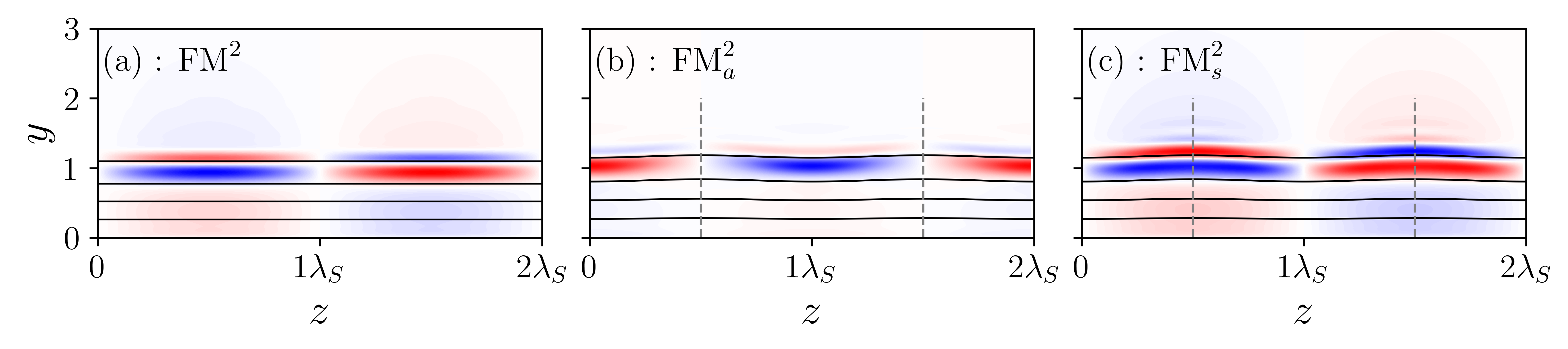}}
    \caption{Velocity fields $u'_x$ of the unstable LST-2D eigenfunctions for $f=0.1$}
    \label{fig:2d_lst_FM_x1_x20}
\end{figure}

In Fig.~\ref{fig:2d_lst_x1_x20}a, for $x=25$ two unstable eigenvalues arise close to the $\text{FM}^2$ mode at $x=0$. Their eigenfunctions shown in Fig.~\ref{fig:2d_lst_FM_x1_x20}b \&~\ref{fig:2d_lst_FM_x1_x20}c, show that these two modes can be understood as symmetric and antisymmetric FM modes, which are degenerate with the streak amplitude going to zero, i.e., they converge to the same eigenvalue. The symmetry imposed by the streaks results in a splitting of the FM instability into two independent instabilities. Similar symmetric and anti-symmetric first-modes were found for $M_\infty=3$ flows by \citet{paredesInstabilityWaveStreak2017}, It is here verified that such modes are indeed connected to the original first mode instability.

\begin{figure}[t]
    \centering
    \centerline{\includegraphics[width=\textwidth]{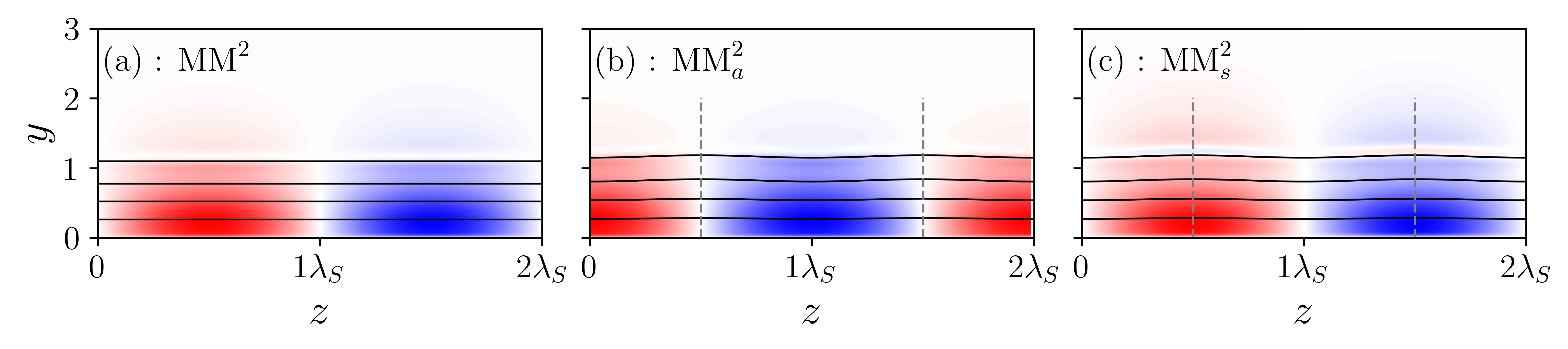}}
    \caption{Velocity fields $u'_x$ of the unstable LST-2D eigenfunctions for $f=0.32$}
    \label{fig:2d_lst_MM_x1_x20}
\end{figure}

The same analysis carried out for the MM instabilities at a frequency $f=0.32$ in Fig.~\ref{fig:2d_lst_MM_x1_x20} reveals two unstable eigenvalues (red) for the undisturbed boundary layer. These two eigenvalues are identified as the planar second-mode instability $\text{MM}^0$ and an oblique second-mode instability $\text{MM}^2$. Adding a slight distortion leads to the blue spectrum. This spectrum has three unstable eigenvalues for the streaky baseflow. In the same way as for the FM modes, the oblique mode $\text{MM}^2$ splits into symmetric and antisymmetric components, respectively $\text{MM}^2_s$ \& $\text{MM}^2_a$. 

The investigation of the streaky hypersonic boundary layer stability in the limit of small streaks has revealed the appearance of new instabilities. In the next section, these unstable modes are further studied as the streaks grow along the domain.

\subsection{Streamwise evolution of convective instability characteristics}

Using the 2D local framework, the spatial spectrum is computed for 30 successive streamwise stations along the baseflow to track the various instabilities. This analysis characterises the effect of streaks on the stability of FM, SM and MM modes in hypersonic boundary layers, and it allows us to complete the characterisation that will be used to support and discuss the global LDNS analysis to follow.

For each baseflow slice $\bar{\pmb q}_{S}^i(y,z),\ i\in[1,30]$, a spatial stability problem is solved. The solver is set to look for the $N_\alpha=200$ eigenvalues closest to the target $\tau$. The unstable eigenvalues detected at $x=0$ or $x=350$ are tracked by recursively projecting their eigenfunctions $\psi_{i,j},\ j\in[1,N_\alpha]$ for the slice $i$ onto the basis of eigenfunctions of the slice $i+1$. Therefore, the eigenfunction maximising $\langle \psi_{i,j}, \psi_{i+1,j} \rangle$ at $i+1$ gives the next position of the tracked eigenvalue in the complex plane as the flow evolves downstream. In practice, a projection coefficient of $\langle \psi_{i,j}, \psi_{i+1,j} \rangle \geq 0.98$ was observed between two successive eigenfunctions of the same eigenvalue. 

\begin{figure}
    \centering
    \centerline{\includegraphics[width=1\textwidth]{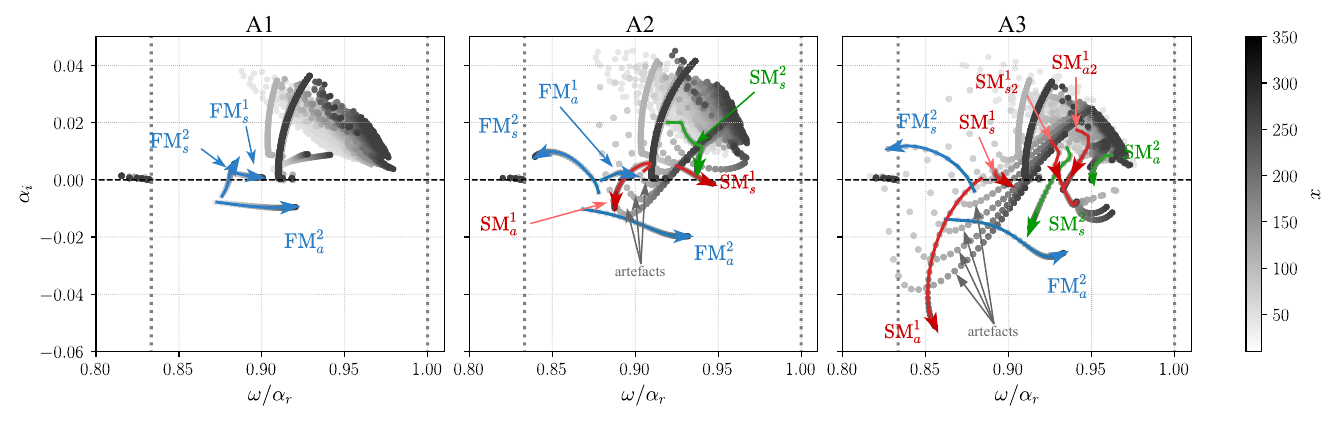}}
    \caption{Evolution of the stability of FM and SM waves as a function of streaks amplitude $A_{s}$ (greyscale) for $f=0.09$.}
    \label{fig:2d_lst_spectrum_compare}
\end{figure}

A first analysis is performed around $f=0.1$ to track the FM and the SM instabilities. The results are shown in Fig.~\ref{fig:2d_lst_spectrum_compare} for the cases A1, A2 \& A3. On these spectra, the streamwise evolution of the eigenvalues is visualised using a gray scale progression from white at $x=0$ to black at $x=350$. Tracked unstable eigenvalues are highlighted in colour. Specifically, FM instabilities are shown with blue arrows, harmonic $\text{SM}^1$ instabilities are shown in red and subharmonic $\text{SM}^2$ instabilities are shown in green. An increase in the number of unstable modes occurs with increasing streak strength from A1 to A3. And, the growth rate of the leading unstable eigenvalue is also increasing by factor of five between the A1 and A3 cases. This increased number of unstable FM and SM modes and their growth rate increase emphasizes an overall destabilizing effect of streaks for convective instabilities at $f=0.1$. This is contrary to the effect of streaks in incompressible flows where they have a globally stabilising effect, as shown by \citet{cossuStabilizationTollmienSchlichting2002}

Looking at the details of Fig.~\ref{fig:2d_lst_spectrum_compare}, some other interesting behaviours of the FM family emerges. From tracking the three modes $\text{FM}^1_s$, $\text{FM}^2_s$ \& $\text{FM}^2_a$ we note that the $\text{FM}^2_s$ is largely stabilized by the streak and its phase speed converges toward the slow acoustic branch. The $\text{FM}^2_a$ mode, on the other hand, remains unstable for all streak amplitudes and is progressively destabilised as the boundary layer distortion becomes more marked. Finally, even if some slight destabilisation effect can be noted, the streaks have a weak effect on the instability of the fundamental $\text{FM}_{s/a}^1$ modes which remain in the stable half-plane. This behaviour of the fundamental FM modes differs from what was observed by \citet{paredesInstabilityWaveStreak2017} where the fundamental FM mode was found to be unstable in a supersonic streaky boundary layer at $M_\infty=3.0$. This suggests an interesting Mach number effect that might be explored in a future study.

SM instabilities are also observed, and these become unstable for the case A2 on Fig.~\ref{fig:2d_lst_spectrum_compare}. Two harmonic instabilities $\text{SM}^1_a$ and $\text{SM}^1_s$ can be seen to originate from the vorticity entropy branch and become progressively more unstable as the streaks grow. The independent appearance of an unstable $\SM{1}{a}$ wave also differs from the behaviour found at lower Mach number by \citet{paredesInstabilityWaveStreak2017} where it was suggested that $\SM{1}{a}$ modes are the continuation of $\FM{1}{a}$ modes as streak amplitude increases. Tracking the trajectories of the modes in the complexe-plane reveals that the $\FM{1}{a}$ mode in fact has its origin in the acoustic continuous branch \citep{fedorovTransitionStabilityHighSpeed2011} whereas the $\SM{1}{a}$ mode originates from the vorticity-entropy continuous branch. Following previous studies on receptivity from local stability theory \citep{fedorovPrehistoryInstabilityHypersonic2001,maReceptivitySupersonicBoundary2003}, where the receptivity process is discussed based on the initial continuous branch from which a discrete eigenvalues emerges. The entropy/vorticity origins of the $SM$-waves compared to the acoustic origin of the $FM$ and $MM$ eigenvalues indicates a different nature for the underlying receptivity process producing the $SM$ instability.

This destabilisation of the SM modes is enhanced as streak amplitude becomes strong, as in the case A3. Specifically, the sinuous streak-dependant mode $\text{SM}^1_a$ rapidly becomes the leading instability mechanism ahead of the sinuous FM instability. Furthermore, the A3 case exhibits a rich ensemble of linear-instability possibilities. In addition to the harmonic SM modes, subharmonic SM instabilities also arise, highlighted in green. The $\text{SM}^2_s$ instability is the third most unstable eigenvalue and its downward trajectory in the complex plane indicates that it is also strongly destabilised by the streak growth.

\begin{figure}
    \centering
    \includegraphics[width=0.8\textwidth]{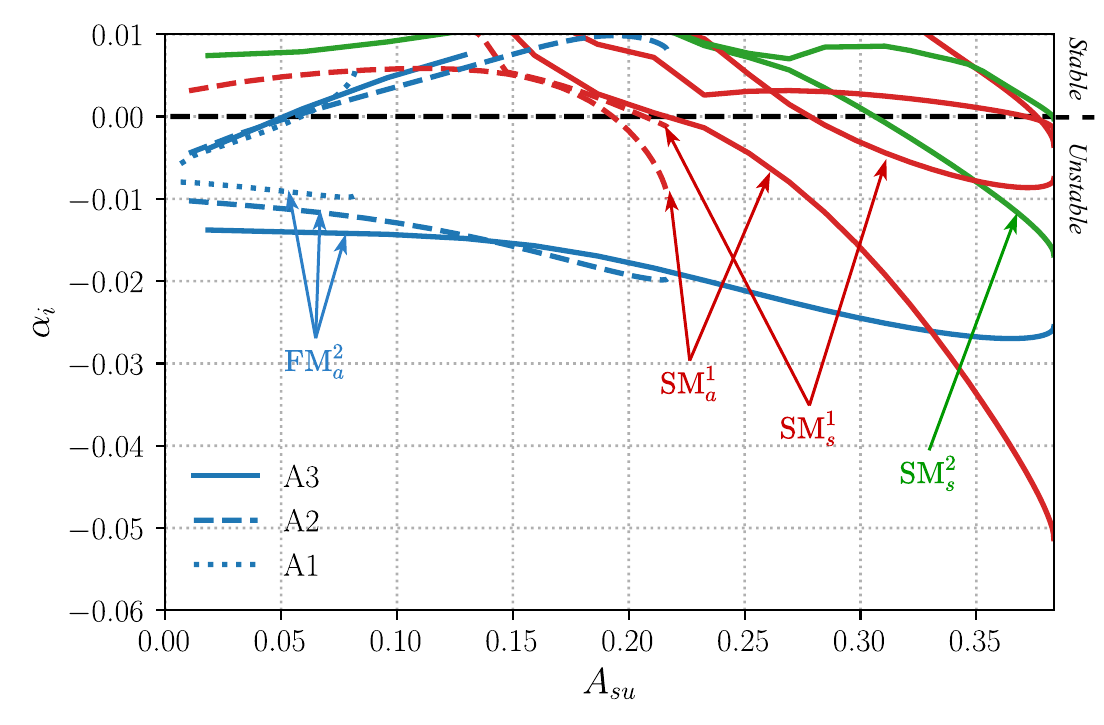}
    \caption{Identified unstable modes growth rate evolution with streak amplitude for the three cases considered. Lines end at the maximum streaks amplitude attained for each baseflow. The colour code used to identify the modes is consistent with Fig.~\ref{fig:2d_lst_spectrum_compare}}
    \label{fig:stk_unstable_amp}
\end{figure}

Better insights on the dominance of each instabilities can be gained by plotting the spatial growthrate $\alpha_i$ as a function of $A_{s}(x)$. The result is shown in Fig.~\ref{fig:stk_unstable_amp} for the unstable modes at $f=0.1$. The FM family is seen to dominate the linear dynamics up to $A_s(x)=0.32$ for these flow conditions. Then, the fast increase of the $\text{SM}^1_a$ growth rate leads to a dominance of this instability by the end of the domain. For cases A2 and A3, the $\text{SM}^1_a$ modes are seen to become unstable around $A_{s}(x)\approx 0.2$, which is consistent with the behaviour of SM family found for incompressible flow \citep{anderssonBreakdownBoundaryLayer2001}. The threshold for the SM mode dominance at $A_su(x) \approx 0.32$ is somewhat similar to the value found by \citet{paredesInstabilityWaveStreak2019} for an isothermal wall and therefore confirms a similar instability threshold for an adiabatic wall. 

\section{Linearised DNS of forced streaky boundary layers} \label{sec:ldns}

Having assessed the effect of streaks on boundary layer instability in a locally parallel framework, we now consider the global problem by computing the linear response of the 3D distorted boundary layer to white-noise forcing. This noise is not intended to be representative of real atmospheric or experimental conditions. It allows, rather, an exploratory analysis of the linear flow response to forcing over a broad range of frequencies, and includes the receptivity process without bias towards a specific instability. Hence, no specific frequency band is preferred. Such forcing method has proven an efficient means by which to observe the various amplification mechanisms in transitional flows \citep{lugrinTransitionScenarioHypersonic2021, haderSimulatingNaturalTransition2018, caoStabilityHypersonicFlow2023, unnikrishnanLinearNonlinearTransitional2020}
\subsection{Forcing and sampling}
 
Using the expression introduced in Eq.~\ref{eq:disturb}, a second forcing vector $\pmb f_w$ is built following the procedure introduced by \citet{haderSimulatingNaturalTransition2018}. A spatio-temporal white noise is constructed by introducing a random pressure field at each point of the grid, giving the following definition for the forcing term with the uniform probability distribution $\mathbf{U}$.

\begin{align}
    \pmb f_{w}(i,j,k) = \pmb X(i,j,k), \quad \pmb X \sim \mathbf U \left\{-1,1\right\}    
\end{align}

The forcing is also restricted to a volume $\pmb \phi_f$ close to the inlet. The same volume used for the steady streaks is employed. For the numerical scheme to properly capture the injected energy and propagate it downstream, the white-noise field is updated every 20 iterations of the solver. With this procedure, the excited frequency range largely contains the instabilities of interest, ensuring that the white-noise is indeed forcing all the instabilities discussed in Sec.~\ref{sec:2d-lst}. It should be added, that as the dynamics are linearised, no non-linear dynamics can emerge. Therefore, the choice of the forcing amplitude is arbitrary and has been chosen to be $A_f=10^{-1}$.

The white noise forcing is applied to the three streaky baseflows. After a transient corresponding to convection over twice the domain length, the system reaches a statistically stationary state. A sampling of the unsteady dynamics is then performed in time for the full domain volume. The sampling rate is determined based on the instability of the highest frequency to be captured. For this flow, it has been chosen to be the first frequency harmonic of the second Mack mode. Knowing that the highest frequency attained by the MM family for this flat plate flow at the inlet conditions is $f_{MM}=0.2975$, its first harmonic would have a frequency of $2f_{MM}=0.595$. Therefore, to satisfy the Nyquist criterion, the sampling frequency is chosen as $f_s = 1.3$. The flow is sampled for $1300$ convective times $t_c = \delta_0/u_\infty$, resulting in $N_t = 1300$ domain snapshots, leading to a state vector constituted of the successive discrete solutions $\pmb q'_i, \  i \in [1, N_t]$ for the full domain,
\begin{align}
    \pmb{Q} = [\pmb q'_1,\pmb q'_2, \ldots, \pmb q'_{N_t}] \in \mathbb{R}^{N_{pts} \times N_v \times N_t}.
\end{align}
With these parameters, the minimal resolved frequency is $f_{\text{min}} = 0.001$ and the maximum resolved frequency is $f_{\text{max}} = 0.65$.

\subsection{Disturbance fields}


\begin{figure}
    \centering
    \begin{subfigure}[t]{0.32\textwidth}
        \includegraphics[width=\textwidth]{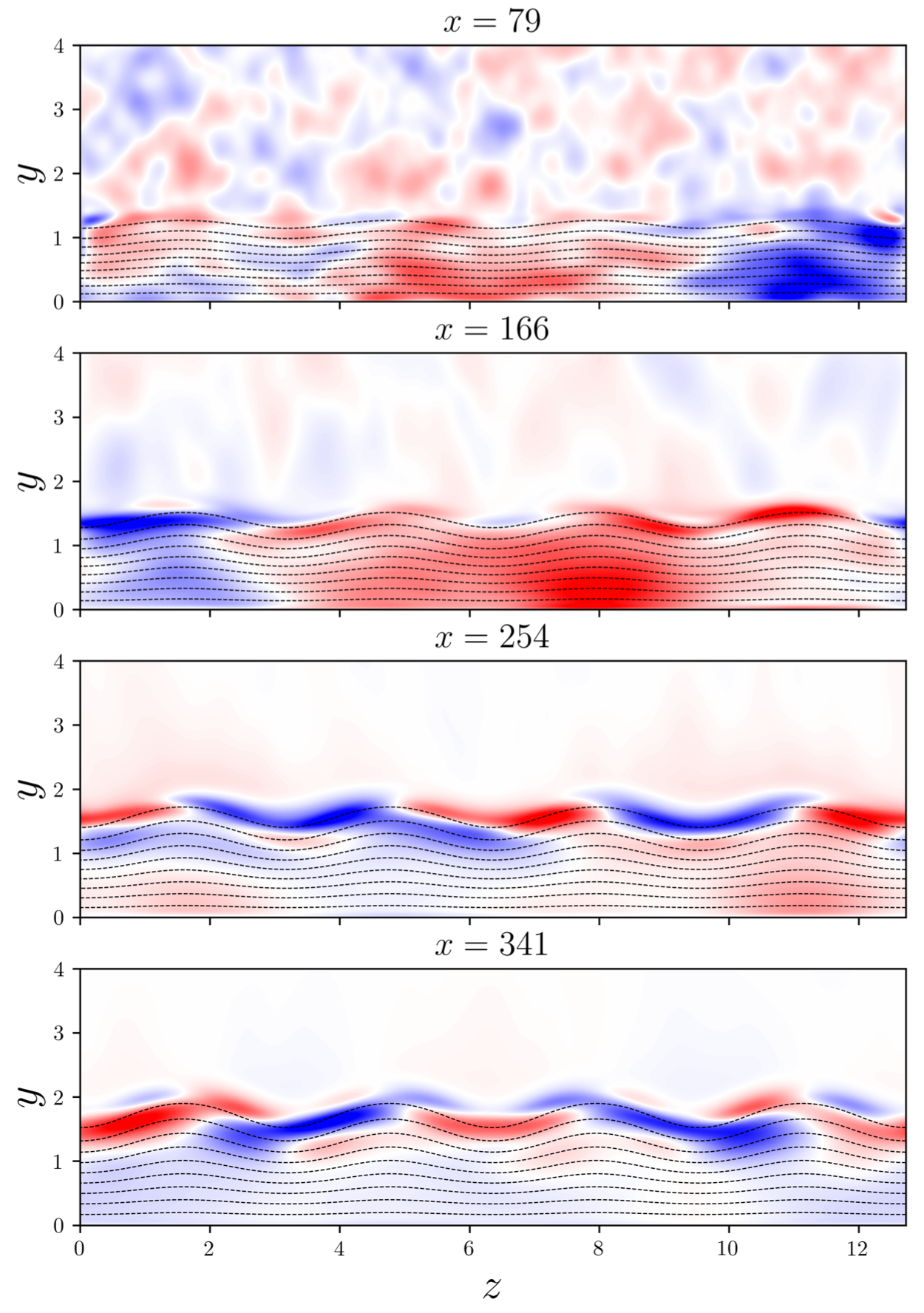}
        \caption{A1 case}
        \label{fig:dist_ux1}
    \end{subfigure}
    \hfill
    \begin{subfigure}[t]{0.32\textwidth}
        \includegraphics[width=\textwidth]{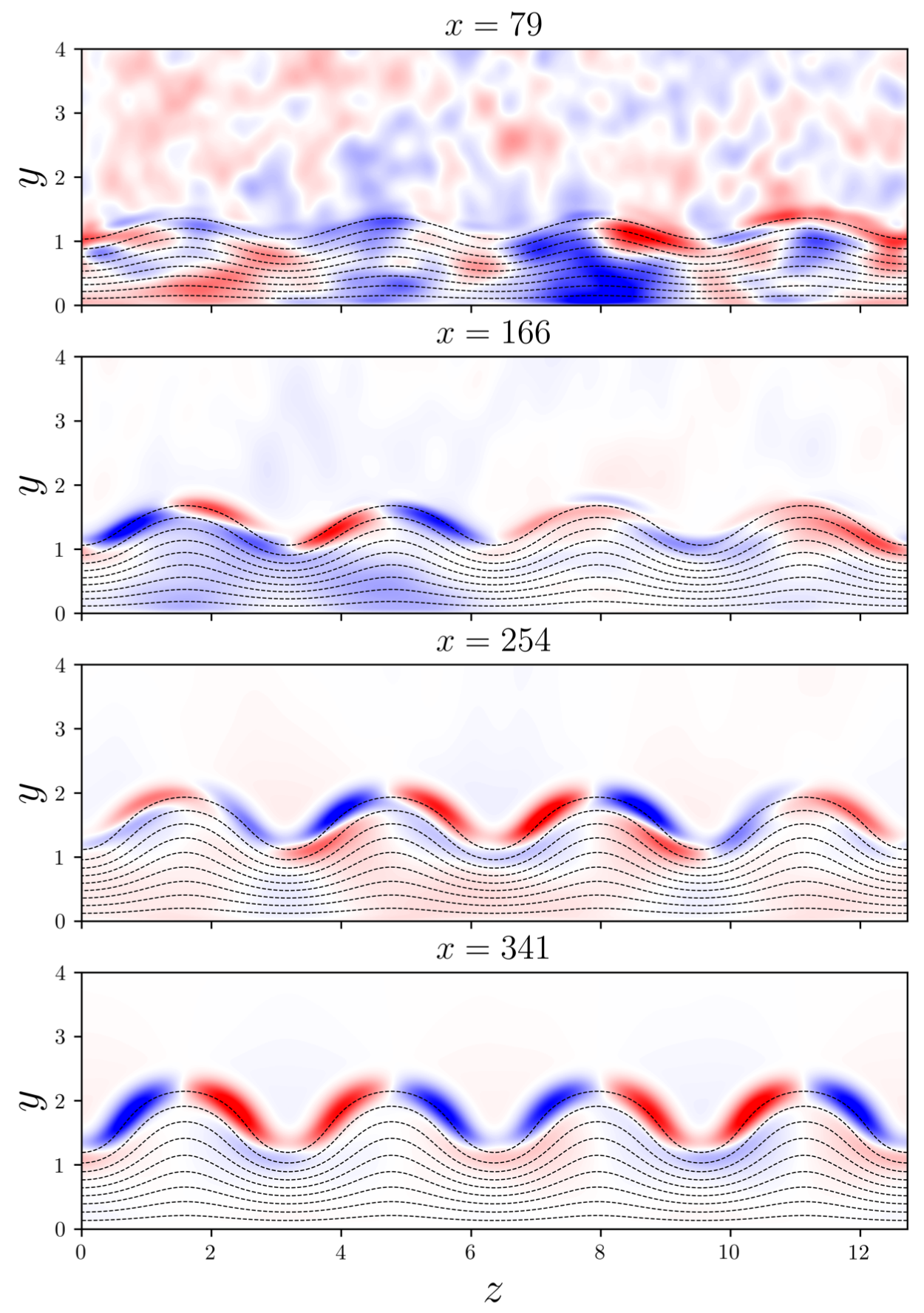}
        \caption{A2 case}
        \label{fig:dist_ux2}
    \end{subfigure}
    \hfill
    \begin{subfigure}[t]{0.32\textwidth}
        \includegraphics[width=\textwidth]{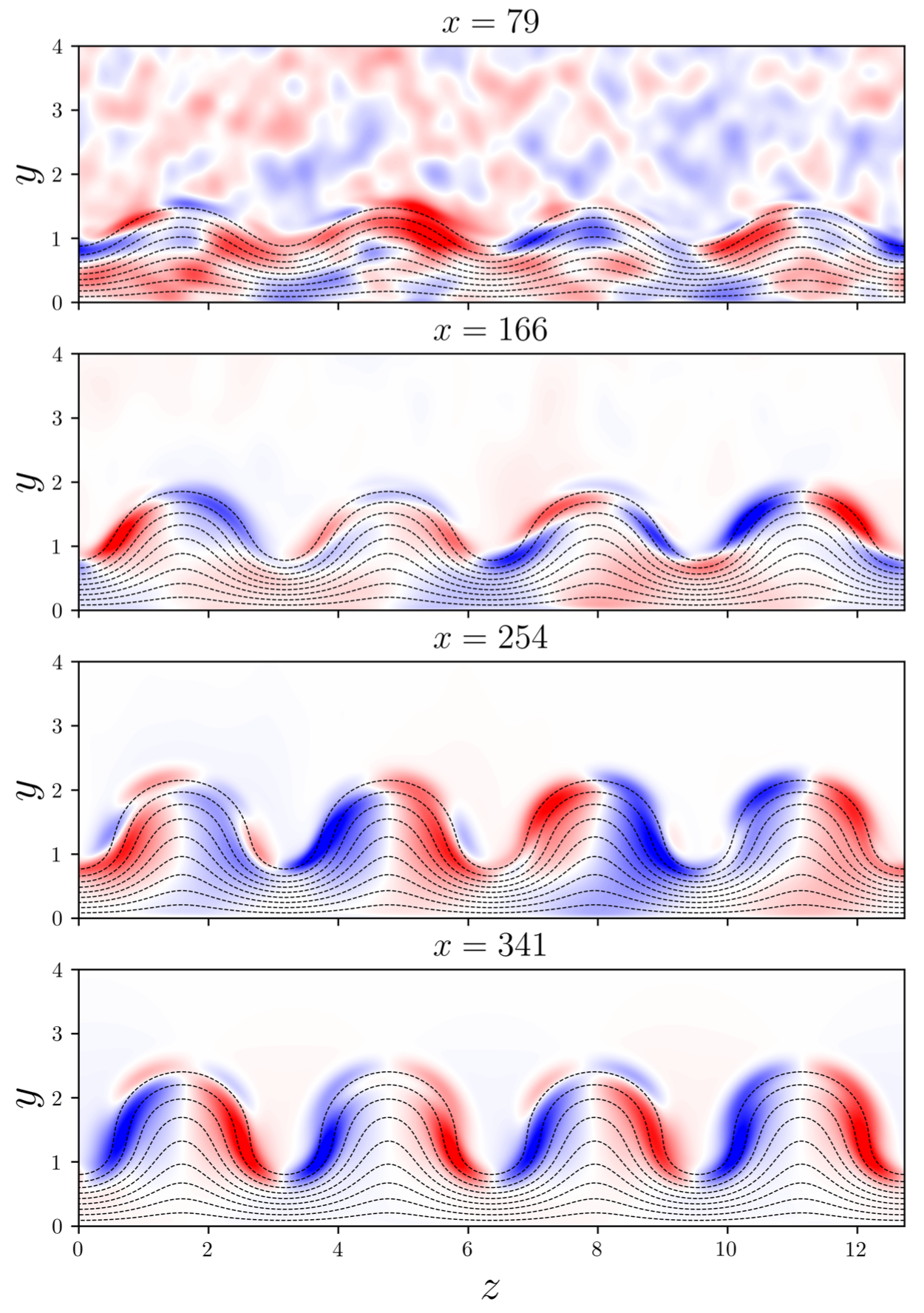}
        \caption{A3 case}
        \label{fig:dist_ux3}
    \end{subfigure}
    \caption{Axial velocity disturbances $u_x'$ over the three baseflows. Contours are normalised with $u_x' \in [-1, 1]$ for each slice. Dashed lines : ten contours of baseflow velocity $u_x\in[0,0.95u_\infty]$}
    \label{fig:dist_ux}
\end{figure}
A first overview of the white noise effect is provided by looking at the snapshots of the disturbance fields evolving around the three streaky boundary-layers. For each case A1, A2 and A3, the axial velocity disturbances $u'_x$ at different streamwise stations are shown in Fig.~\ref{fig:dist_ux}. The white noise forcing is visible at $x=79$ for all three cases. Within the boundary layers larger flow structures can be seen, suggesting that receptivity and preliminary amplification has taken place. Further downstream, at $x=166$, linearly amplified flow structures are observed. At this position, the structures already differ between the three cases. Specifically case A1 shows larger structures spanning multiple streaks and that extend over the entire boundary layer height. On the other hand cases A2 and A3 show more compact structures, mostly located at the upper edges of the streaks. Looking at the next station, $x=254$ such localised structures become more noticeable in all cases. Finally, at the end of the domain $(x=343)$, a clear pattern is visible for case A2 with what looks like a subharmonic organisation.  For cases A1 and A3, the disturbance fields yields a less clear conclusion as different distributions of various spanwise periodicity look to be superposed. In what follows, further decomposition and comparison with local stability characteristics will clarify these structures.

\subsection{$N-$periodic flow decomposition}

In order to perform quantitative identification of the flow structures observed in the disturbances field, a signal decomposition is carried in the spectral space \citep{tairaModalAnalysisFluid2017}. This decomposition of the state vector $\pmb q'(x,y,z,t)$ is done on the frequency and spanwise-wavenumber dimensions to obtain the linear response to white-noise in the form of $\pmb q'(x,y,\beta,f)$. 

With data in the frequency-wavenumber domain, the Spectral Proper Orthogonal Decomposition (SPOD) is now commonly used to decompose state vectors at independent frequency into a basis of orthogonal modes ranked by energy with respect to a given norm \citep{lumleyStochasticToolsTurbulence1970, schmidtGuideSpectralProper2020}. However, for N-periodic baseflows, difficulties arise in obtaining a computationally efficient decomposition. An extension of the classical SPOD formulation to take advantage of the N-periodicity of the baseflow using Floquet expansion (or also Bloch expansion) has been specifically developed for this work and is presented in Appendix~\ref{sec:appendix-a}. 


\begin{figure}
    \centering
    \centerline{\includegraphics[width=0.9\textwidth]{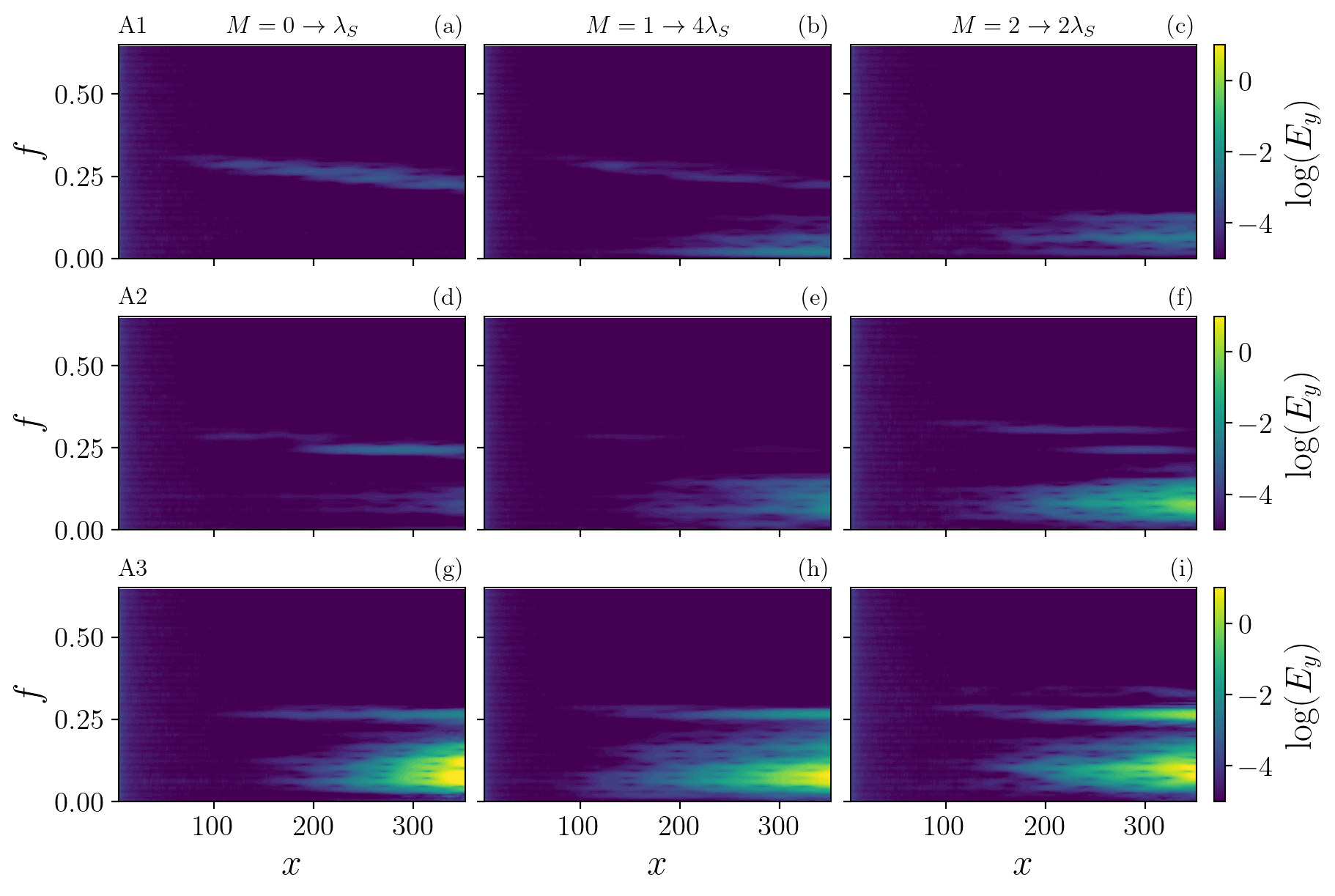}}
    \caption{Streamwise energy evolution of the frequency spectrum. From top to bottom, cases A1, A2 and A3. Columns titles describe the wave periodicity and the corresponding Floquet exponent as defined in Appendix \ref{sec:appendix-a}}
    \label{fig:lin_spectrum_x}
\end{figure}
In order to measure the amplification of the structures observed in Fig.~\ref{fig:dist_ux}, the inner product formulation introduced in Appendix~\ref{sec:appendix-a} (Eq.~\ref{eq:inner-floq}) is used to compute the energy of disturbances integrated over the boundary layer height at each frequency and streaks periodicity.
\begin{align}
    E_y = \frac{1}{\lambda_z}\int_y \int_z \left\langle \tilde{\bm{\mathsf X}}_{\omega_i,\gamma_M}, \tilde{\bm{\mathsf X}}_{\omega_i,\gamma_M} \right\rangle_W \text{d}z\text{d}y. 
\end{align}
The streamwise evolution of this measure is shown in Fig.~\ref{fig:lin_spectrum_x} for waves of periodicities $\lambda_z \in [\lambda_S, 2\lambda_S, 4\lambda_S]$. Each of these periodicities respectively correspond to a given Floquet exponent $M\in[0, 2, 1]$ in the introduced spanwise decomposition. These spectra show the amplification of different modes as the streak amplitude evolves. For all cases, instability wave growth can be observed from around $x \approx 90$. Two main waves families are visible in all three cases. A first family is related to low-frequency instabilities around $f=0.1$ and a second family to high frequency at $f\approx 0.26$. These frequency intervals are in agreement with the unstable frequencies found with the spatial stability analysis for FM, SM and MM unstable waves. 

An interesting behaviour of the observed instabilities is that the most amplified peaks are quite different from cases A1 to A3. Destabilizing and stabilizing effects can be noted by looking at how the high frequency peaks fade from A1 to A3. On the other hand, low frequency instability is enhanced by the baseflow distortion. 

However, the effect associated with the spatial evolution of the streak amplitude is more subtle when one considers how the different peaks vary from one streak-periodicity to another as streaks grows. For instance, even if the high frequency peaks seem to fade at the harmonic wavelength from A1 to A2 low frequency waves are amplified at this same wavelength. Furthermore, a high frequency peak is seen to destabilise from A1 to A3 at subharmonic wavelengths. Additionally, this peak seems to be subdividing into two peaks as it can be remarked in Fig.~\ref{fig:lin_spectrum_x}f \&~\ref{fig:lin_spectrum_x}i. These behaviours can now be clarified using the nomenclature derived from the local analysis, and the comparison of the local eigenfunctions with the SPOD modes computed from the data

\subsection{SPOD of the boundary-layers responses at $x=350$}

%
\begin{figure}
    \centering
    \includegraphics[width=\textwidth]{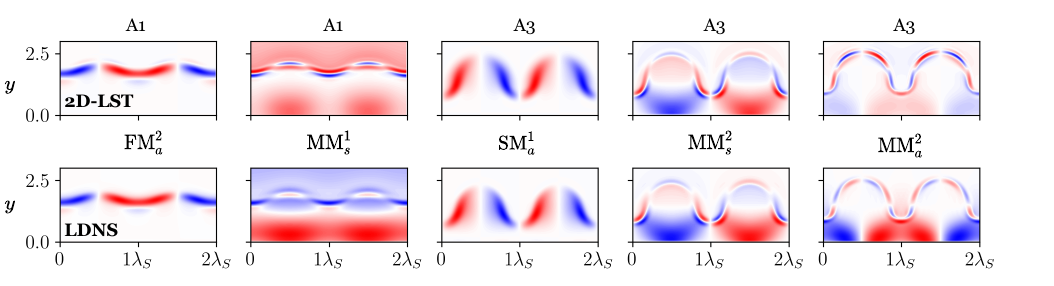}
    \caption{SPOD / 2D LST modes comparison }
    \label{fig:spod_dns}
\end{figure}

\begin{figure}
    \centering
    \includegraphics[width=\textwidth]{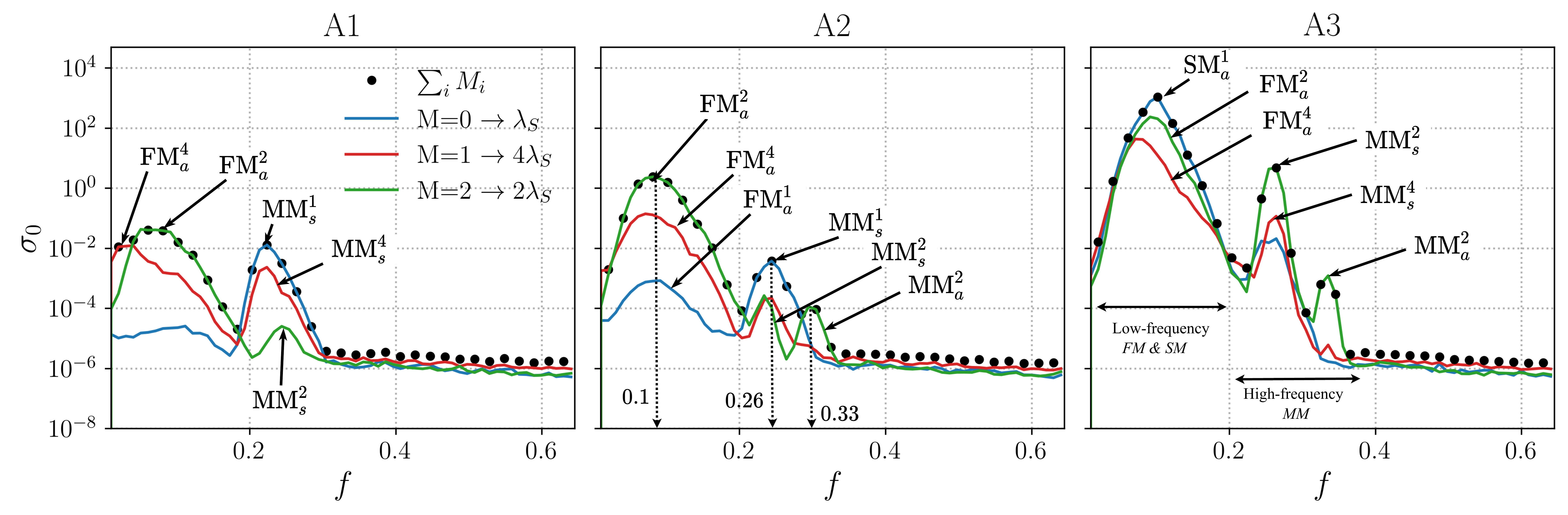}
    \caption{LDNS SPOD energy of the leading mode $\sigma_0$ at end of domain for the three cases }
    \label{fig:floquet_spod_gains}
\end{figure}

Floquet-SPOD is applied at the end-plane of the domain for baseflows A1, A2 \& A3. The resulting of the leading SPOD mode are shown in Fig.~\ref{fig:floquet_spod_gains}. On each plot, the energy spectrum is split into its three Floquet periodicities. The dotted line shows the total energy of the full spanwise spectrum whereas the blue, green and red lines correspond, respectively, to the energy levels of the harmonic, subharmonic and $\lambda_z=4\lambda_S$ wavelengths. Given the comparisons between SPOD modes and 2D eigenfunctions, shown in Fig.~\ref{fig:spod_dns}, the Fig.~\ref{fig:floquet_spod_gains} permits an identification of the linear instabilities that dominate by the end of the domain, and how streak amplitude affects these.
%

%
For the low-amplitude streak, case A1, the dominant peak is at low frequency, $f = 0.09$ and corresponds to a subharmonic antisymmetric first-mode instability ($\text{FM}^2_a$). A second peak is seen at high frequency for an harmonic wave corresponding to a second-mode instability $(MM^1_s)$ at $f=0.24$. For the $MM$ waves, the subharmonic component is only weakly amplified whereas, for $\lambda_z = 4\lambda_S$, a substantial amplification of low frequency and high frequency component can be seen. The amplified waves observed at $x=350$ thus shows that transition may involve multiple linear instabilities, even for the low amplitude streaks of A1 ($A_{s}<0.1$).

As the streak amplitude is increased to $A_{s}\approx 0.2$ for the A2 baseflow, the low frequency $\text{FM}^2_a$ mode has undergone larger growth, and dominates by a factor $10^3$ the $\text{MM}^1_s$ mode. This latter instability has undergone less growth (by a factor of $10$) in comparison to the A1 case. And, the $\text{FM}^4_s$ has also undergone more growth than in the A1 case. It can then be argued that this streak amplitude is favouring the growth of multiple, low frequency, FM instabilities and which may collectively underpin transition.

At the end of the domain for the case A2, the $\text{MM}^2_{a/s}$ instabilities are found to split into antisymmetric and symmetric components, 
visible in green. These peaks now correspond to the modes $\text{MM}^2_s$ and $\text{MM}^2_a$ visible as two high-frequency humps, shown in Fig.~\ref{fig:spod_dns}. their spatial support highlights their trapped-acoustic-wave nature \citep{fedorovTransitionStabilityHighSpeed2011}. Knowing that the MM wave frequency is inversely proportional to the boundary layer height $(f_{MM} \propto  2 u_\infty / \delta_{99})$, the boundary layer distortion induced by the streaks may significantly alter the second-mode frequency. Specifically, two frequency bounds can be defined at a given streamwise station by taking the lowest and highest point of the boundary layer at, respectively, the high-speed and low-speed streaks as depicted in Fig.~\ref{fig:stk-bf}. Therefore, the symmetric $\text{MM}^2_s$ mode being mostly located  under the low speed streaks, corresponding to the highest point in the boundary layer, have the lowest frequency. On the other hand the antisymmetric $\text{MM}^2_a$ mode which is concentrated under the high-speed streaks of lowest boundary layer height, presents a higher frequency. This observation is further confirmed by looking at the proportionality relation between the two mode frequencies and the lowest/highest boundary layer points. In the case of A2 at $x=350$, it gives $f^\text{min}_\text{MM} / f^\text{max}_\text{MM} = \delta_{99}^{\text{min}}/\delta_{99}^{\text{max}}=0.76$. Such a frequency separation of the second-mode waves has not been clearly identified in previous studies to the best of our knowledge and may offer additional insights in the interpretation of numerical or experimental data for distorted boundary layers , i.e., they support two types of Mack modes, each one driven by the effective BL thickness found on slow and fast streaks

Increasing further the streak amplitude up to $A_{s}\approx 0.4$ for the A3 case leads again to substantial changes in the dominant instabilities. The low frequency waves see their amplitude increased by a factor $10^2$ in comparison to A2. Within the low-frequency waves, three distinct peaks with close amplitudes are observed. These peaks are now dominated by a $\text{SM}^1_a$ instability, meaning that the streaks have reached a sufficiently high amplitude for these intrinsic streak-dependent instabilities to grow. The two next-most-amplified modes ranked by decreasing energy are the $\FM{2}{a}$ and $\FM{4}{s}$. The close amplitude levels of these instabilities suggests that they may all play a considerable role in the route to turbulence.

At higher frequencies, the $\MM{4}{s}$ and $\MM{2}{a}$ second-mode waves are amplified but the $\MM{1}{s}$ wave remains at a similar amplitude. Specifically, the subharmonic $\MM{2}{s}$ mode has the highest amplitude, but remain $10^2$ times smaller than the $\SM{1}{a}$ mode. Nonetheless, these second-mode waves are amplified by a factor $10^5$ in comparison to the A2 baseflow which implies a strong enhancement of their growth by the A3 streaks. Additionally, the frequency separation of the symmetric and anti-symmetric second-modes is even more visible as the baseflow distortion is enhanced, consistent the interpretation made above. However, this higher frequency $\MM{2}{a}$ mode is less amplified and may play a secondary role in the transition process in comparison to the other instabilities. 

To summarise the new results observed in this section, the second-mode growth is shown to be non-monotonic with increasing streaks amplitude. Additionally, the leading second-mode switches from a fundamental to a subharmonic wave as the streak amplitude increases. Such behaviour were not discussed previously in the literature. For the first-mode waves, our results show that, similar to what was discussed by \citet{paredesTransitionDueStreamwise2016} for an isothermal boundary-layer, the subharmonic instability is preferred by the adiabatic boundary-layer and is consistently destabilised for increasing streak amplitude. As a new result, we furthermore demonstrate that the first-mode mechanism has a final amplitude dominating the linear dynamics in every case. This finding shows that in presence of streaks, the adiabatic boundary layer gradually switches from second-mode dominated dynamics to first-mode dominated dynamics as streak amplitude increase. This latter point strongly differs from the Blasius case and has important practical implications for transition modelling on realistic walls.

\subsection{Tracking the dominant instabilities}

\begin{figure}
    \centering
    \centerline{\includegraphics[width=\textwidth]{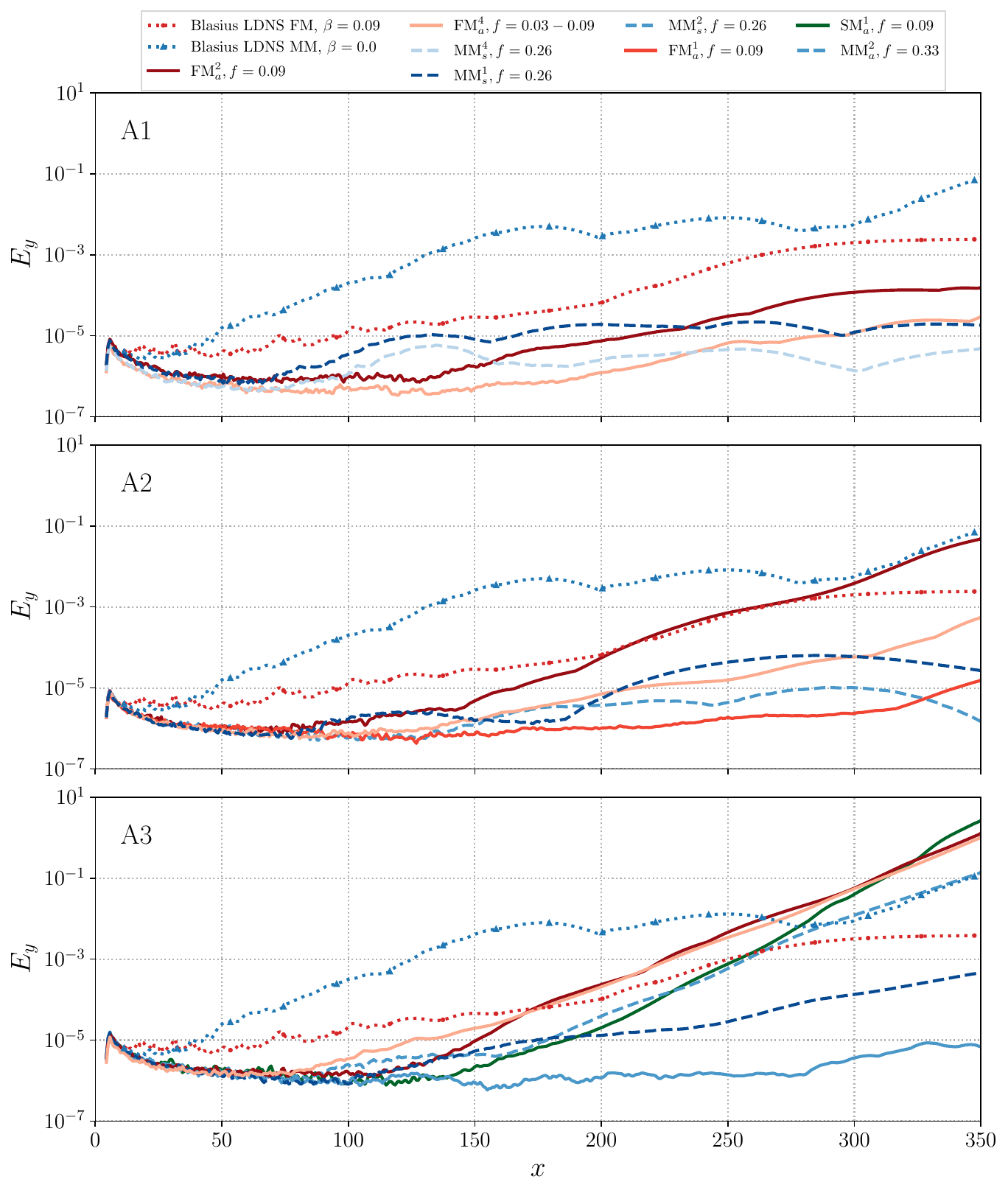}}
    \caption{Streamwise energy evolution of the leading modes of the frequency spectra for each baseflow. Dot lines and markers: dominating modes of the homogenous boundary layer for reference.}
    \label{fig:A5_M2_stream_floq}
\end{figure}

We now track the amplitude evolution the selected instabilities along the domain, the selection is made based on the dominant peaks found in Fig.~\ref{fig:floquet_spod_gains}. These modes are compared to the dominating instabilities observed for the baseline case of a Blasius flat-plate LDNS, namely the planar second Mack mode and the oblique first mode. This quantitative comparison between a streaky and non-streaky baseflow gives a useful picture of the effect of streaks on the growth of linear instabilities. The comparison is shown in Fig.~\ref{fig:A5_M2_stream_floq} for cases A1, A2 \& A3.

The first- and second-mode instabilities of the Blasius boundary layer are plotted using dotted lines with markers. As expected from local analysis, the second Mack mode dominates by at least an order of magnitude. As the wall is adiabatic, the first mode also undergoes substantial amplification. Therefore, in such a case one would argue that the second-mode should dominate the transition to turbulence. 

The effect of streaks is visible starting from the A1 case in Fig.~\ref{fig:A5_M2_stream_floq}. For this baseflow, the instabilities corresponding to the main peaks identified in Fig.~\ref{fig:floquet_spod_gains}, $\FM{2}{a}$, $\FM{4}{a}$, $\MM{1}{a}$ \& $\MM{4}{a}$ show amplitude levels well below the first and second modes of the Blasius flow. These results imply a strong stabilizing effect of the low-amplitude streaks on both FM and MM families of instability. But looking closely at the amplification of the leading FM mode in case A1, the spatial growth is of similar magnitude to the A0 casecase, but shifted downstream. This suggest a change in receptivity mechanism. Another difference with the Blasius case is that the MM mode cannot be said to be dominant, as at the end of the domain it is the FM modes that achieve largest amplitudes. In light of recent analysis in the stabilizing properties of streaks in hypersonic flows \citep{paredesInstabilityWaveStreak2019}, the results of case A1 indicates similar stabilizing properties on MM modes.

For the A2 case, in the upstream part of the boundary layer, a global stabilisation is again observed. But towards the end of the domain, the $\FM{2}{a}$ instability achieves amplitudes larger than that of its Blasius counterpart and comparable to that of the most unstable Blasius Mack mode. This suggests that for moderate-amplitude streaks, the key mode underpinning transition has changed from MM to FM, and the streaks can not be concluded to have had a global stabilising effect. A downstream shift is again observed in the initiation of instability wave growth, pointing to a change in receptivity mechanisms as the for eh A1 case. But in this case the spatial growth rates in the streaky base flow are larger than in the Blasius case. The moderate-amplitude streaks can thus be concluded to produce an overall destabilisation.


In the A3 baseflow, the $\FM{2}{a}$ and $\FM{4}{a}$ instabilities are further destabilised, reaching final amplitude $10^2$ greater than MM for the A0 baseflow. These modes can be expected to play an important role in the transition process along with the streak instability $\SM{1}{a}$ which undergoes substantial growth in the downstream section of the boundary layer. Indeed, it overtakes the FM instabilities around $x \approx 325$ and dominates the linear dynamics at the end of the domain. As for the MM instabilities, they all show substantial growth throughout the domain. The $\MM{1}{s}$ instability remains of lower amplitude, while, the $\MM{2}{s}$ mode shows a growth rate close to the FM and SM instabilities. These trends indicate that for these high-amplitude streaks, the baseflow supports a rich ensemble of linear instabilities that provide multiple paths to turbulence. A summary of these stabilising and destabilising effects for the different observed instabilities is given in Tab.~\ref{tab:stk_stab}

\begin{table}
    \centering
    \begin{tabular}{c|cccc}
           & $\text{FM}^2_a$ & $\text{MM}^1_s$ & $\text{MM}^2_s$ & $\text{SM}^1_a$ \\\hline\hline
        A1 & Destabilizing & Stabilizing & Stabilizing & stable \\
        A2 & Destabilizing & Stabilizing & Stabilizing & stable \\
        A3 & Destabilizing & Destabilizing & Destabilizing & Destabilizing\\
    \end{tabular}
    \caption{Qualitative summary of growing streaks effect on identified instabilities}
    \label{tab:stk_stab}
\end{table}

Finally, we note that for all instabilities tracked in Fig.~\ref{fig:A5_M2_stream_floq}, the receptivity regions defined by the distance between the noise injection and the first growth was seen to be considerably larger than for the non-streaky case (approximately $100\delta^*$ longer). The noise generating procedure being strictly identical between the streaky and non-streaky simulations, this suggests an effect of the streaks in the receptivity process, which seems to exhibit a more extended streamwise support. This mechanism is not fully understood yet but the induced vorticity generated by the optimal forcing might play a role in changing the injected noise projection on the basis of optimal-forcing. This is to be the subject of future studies. 

\section{Conclusion} \label{sec:conclusion}

This study sought to explore the impact of streaks on linear instability mechansisms that arise in wall-bounded hypersonic flow,  and, in particular, for cases involving the growth of steady streaks. A review of prior research on streak instabilities in both incompressible and compressible flows revealed a knowledge gap regarding the typology and behaviour of instabilities in streaks-containing hypersonic boundary layers.  

To address this knowledge gap, we undertook a study based on based on a time marching of the 3D linearized Navier-Stokes equations. This allowed us to explore the linear response of the streaky boundary layer to white-noise forcing and thereby study the numerous instability mechanisms that may arise. These simulations were complemented by 2D local stability analyses at various streamwise stations. The stability analysis enabled the identification, classification and tracking of convective instabilities in the complex plane, and their connection to the linear growth mechanisms that exist in the Blasius boundary layer.

Considering the spanwise periodicity of the streaky base-flow structure, particular attention was given to modal decomposition of dynamics. Specifically, we developed a formulation the SPOD framework, tailored to N-periodic compressible baseflows, utilizing Floquet theory. The use of the Floquet formulation yielded substantial cost reductions compared to established methods and provided a clear spatial separation of the coherent structures, which, combined with the locally parallel stability analysis, enabled a clear classification of the various mechanisms observed in the simulations. This methodology is applicable to any N-periodic baseflow.

Decomposition of the LDNS databases for three levels of streak amplitude highlighted the complex effect of streaks on convective instability dynamics. For low-amplitude streaks, the boundary layer shows an overall stabilisation, with delayed onset of the growth for the first- and second-mode instabilities. However, while the growth rate of the second mode is reduced, the growth rate of the first mode increases significantly compared to its Blasius counterpart.

With moderate streak amplitudes, the first-mode instability is further destabilised, eventually reaching amplitudes similar to those observed in the Blasius flow. Increasing the streak amplitude further led to a general destabilisation of the boundary layer, characterised by a rapid rise in first-mode instabilities along with streak-specific instabilities that exceeded the amplitudes observed in the Blasius flow. Last, the second mode was found to undergo a frequency separation, with low- and high-frequency Mack modes, associated with the two characteristic boundary layer heights imposed by the streaks. Importantly, we observed for the first time that this frequency splitting is proportional to the modulation of the boundary layer height, for the hypersonic-flow studied here. If this result is confirmed for other configurations, it can become a useful for modeling MM-waves in streaky flows.

In summary, the study sheds light on the rich diversity of linear dynamics that arise in streaky boundary layer flows, and the numerous paths to transition that may therefore exist. Future research directions may include an analysis of streaky flow receptivity in a global resolvent framework, as introduced by \cite{jouinCollectiveSecondaryInstabilities2022}, and an exploration of the saturation and nonlinear pathways to turbulence of the observed instabilities, contributing further to our understanding of the laminar-to-turbulent transition on baseflows modulated by streaks.


\section*{Acknowledgment}

This work is part of a project supported by Région Nouvelle-Aquitaine under grant 2018-1R10220. This work has been granted access to the HPC resources of IDRIS under the allocations A0092A10868 and A0112A10868 made by GENCI (Grand Equipement National de Calcul Scientifique). 

\appendix 

\section*{Declaration of Interests}
The authors report no conflict of interest.

\section{Floquet-SPOD formulation for N-periodic baseflows} \label{sec:appendix-a}

In what follows, useful details related to building the SPOD over $N-$periodic compressible baseflows are discussed.
%
%

The transformation to Fourier space in the $t$ and $z$ dimensions requires two different approaches. For the time, an estimate of the frequency spectrum is obtained using the classical Welch method with blocks of size $N_{\text{fft}}=128$ and an overlap of $50\%$. Hanning windows are applied on each blocks to avoid spectral leaking due to the signal cut-off. 

For the spanwise direction $z$, the baseflow is inhomogeneous $(\partial \bar{\pmb q}/\partial z \neq 0)$ but $N-$periodic with $N=4$ streaks. Therefore, a decoupling of the flow coherent structures by single spanwise Fourier modes is not possible as the disturbance field is coupled with the streaks shape \citep{anderssonBreakdownBoundaryLayer2001} that are described by a set of Fourier modes as seen from the multiple peaks of Fig.~\ref{fig:stk_bf_fft}. In spite of that, making use of this $N-$periodicity for the inhomogeneous base-flow decomposition if of importance for the computation of the SPOD modes. Indeed, obtaining the 3D SPOD eigenbasis with the full $z$ direction in space can be cumbersome to compute considered the large size of the $\pmb{Q}$ vector ($> 20$To for each baseflow). 

This issue was previously identified by \citet{rigasStreaksCoherentStructures2019} who initially decomposed a serrated jet $N-$periodic flow in the frequency-wavenumber domain in order to select relevant Fourier modes in $z$ but then reverted to the frequency-space domain to perform an SPOD of the disturbances field over the $N$-periodic base state. This choice to revert to the frequency-space domain before computing the SPOD was guided by difficulties in computing the inner product for the baseflow-dependant Chu norm in the wavenumber domain. If one were able to perform the SPOD in the frequency-wavenumber domain, only a reduced amount of modes in $z$ could be used to describe the disturbance field, greatly lowering the cost of obtaining 3D SPOD modes. Therefore, the next section introduce an extension of the SPOD formulation for $N$-periodic basestates in the frequency-wavenumber domain by solving the inner product formulation issue.

\subsection{Spanwise Floquet expansion of the flowfield}

Advantage can be taken from the periodic boundary-conditions and the $N-$periodic nature of the baseflow to use a Floquet expansion of $\pmb q'$ in $z$. Following the expansions previously used by \citet{kopievAeroacousticsSupersonicJet2004,sinhaParabolizedStabilityAnalysis2016,lajusSpatialStabilityAnalysis2019}, the discretized base state is periodic in $z$, with a period $\lambda_s$, therefore it can be written as,
\begin{align}
    \bar{\pmb q}_S(x,y,z) =  \bar{\pmb q}_S(x,y,z + n \lambda_s),\quad n \in \mathbb{N}.
\end{align}
Considering a coupling of the disturbance field to the spatial structure of the baseflow \citep{sinhaParabolizedStabilityAnalysis2016}, the vector $\pmb q'$ can be written as a product of an elementary function $\tilde{\pmb q}'$ periodic in a streak and a complex exponential with a period on $M$ streaks such that, 
\begin{align}
    \pmb q'(x,y,z,t) = \tilde{\pmb q}'(x,y,z,t)\exp(i 2\pi M z / \lambda_s). \label{eq:stk_floquet}
\end{align}
%
%
%
Owing to its $N$-periodicity, the elementary disturbance function $\tilde{\pmb q}'$ of Eq.~\ref{eq:stk_floquet} can also be expanded in a sum of Fourier modes noted $k\in\mathbb{N}$ which are constrained to be multiple of $N$ by the flow topology,
\begin{align}
    \tilde{\pmb q}'(x,y,z,t) = \sum_{k=-N_\zeta/2+1}^{N_\zeta/2}  \hat{\pmb q}'(x,y,t)\exp(-i 2\pi N k z / \lambda_s) \label{eq:stk_floquet_fourier-z}.
\end{align}
Introducing Eq.~\ref{eq:stk_floquet_fourier-z} in Eq.~\ref{eq:stk_floquet}, yields the complete basis of spanwise Fourier modes parametrized by the discrete exponent $\gamma_M = (M - N k)$ on the $N$-periodic baseflow,
\begin{align}
    \pmb q'(t,x,y,z) = \sum_M\sum_{k} \hat{\pmb q}'(x,y,t)\exp(-i2\pi (M - N k) z / \lambda_s)  \label{eq:stk_fourier-z}.
\end{align}
By choosing different values of $M$ the Fourier basis can be separated in different corresponding set of Fourier modes $\gamma_M$,
\begin{align}
    \gamma_M &= \{M - N k \quad | \quad k\in]-N_\zeta/2,N_\zeta/2] ,\ M,N\in\mathbb{N}\}. \label{eq:floquet_set}
\end{align}
Using the same formatting as \citet{rigasStreaksCoherentStructures2019}, an example of such sets is given for the studied case in Tab.~\ref{tab:floq_modes}. 
\begin{table}
    \centering
    $
    \begin{array}{c|ccccccccc}
        \textrm{Mode index in } \gamma_M & 1 & 2 & \ldots & 36 & 37 & 38 & \ldots & N_\zeta/N-1 & N_\zeta/N \\ \hline \hline
        -\frac{N}{2}<M \leq \frac{N}{2} & \multicolumn{8}{|c}{\gamma_M=M-N k} \\ 
        \hline
        -1 & -147 & -143 & \ldots & -5 & -1 & 3 & \ldots & 145 & 149 \\
         0 & -144 & -142 & \ldots & -4 &  0 & 4 & \ldots & 144 & 148 \\
         1 & -149 & -145 & \ldots & -3 &  1 & 5 & \ldots & 143 & 147 \\
         2 & -146 & -142 & \ldots & -2 &  2 & 6 & \ldots & 146 & 150
    \end{array}
    $
    \caption{Example of $\gamma_M$ set found for relevant Floquet exponents of the current baseflow}
    \label{tab:floq_modes}
\end{table}
Each of these set now describe disturbances of $M$-periodicity with respect to a streak, namely harmonic, subharmonic and so on. 

\subsection{Efficient SPOD formulation for $N$-periodic basestates} \label{sec:floq-spod}

The SPOD orthogonal eigenbasis is obtained from a decomposition of the cross-spectral density matrix (CSD). At the core of the CSD matrix computation lies an inner product defined with a given weighting. For hypersonic and more generally compressible flows, a relevant energy measure often used is the norm of \citet{chuEnergyTransferSmall1965} given for each point in Eq.~\ref{eq:chu-norm} with the weight matrix $\bm{\mathsf W}\in\mathbb{R}^{N_v \times N_v}$. 
\begin{align}\label{eq:chu-norm}
    \bm{\mathsf W}=\text{diag}\left(
        \frac{\bar{T}}{\gamma \bar{\rho} M_\infty^{2}}, 
        \bar{\rho}, 
        \bar{\rho}, 
        \bar{\rho}, \frac{\bar{\rho}}{\gamma(\gamma-1) \bar{T} M_\infty^{2}}
       \right)
\end{align}
This choice of energy measure is of importance for the formulation of the SPOD with a Floquet ansatz as it depends on the baseflow structure that is here non-homogeneous. 

The issue is stated by starting from the usual equations obtained for the homogeneous case. In such situation, the SPOD eigenbasis $ \mathbf{\Psi} \mathbf{\Lambda} \mathbf{\Psi}^{-1}$ is obtained through the following decomposition to get the optimal modes at a given frequency-wavenumber duet $(\omega_i,\beta_j)$.
\begin{align}
    \bm{\mathsf S}_{\omega_i} &= \bm{\mathsf {\hat X}}^H_{\omega_i,\beta_j} \bm{\mathsf W} \bm{\mathsf {\hat X}}_{\omega_i,\beta_j}, \label{eq:usual-spod} \\
    \bm{\mathsf S}_{\omega_i} \mathbf{\Psi} &= \mathbf{\Psi} \mathbf{\Lambda}
\end{align}
Where the $\hat{\bm{\mathsf X}}\in\mathbb{C}^{N_{pts}\times N_v}$ is extracted from the frequency-block matrix $\bm{\mathsf{\hat{Q}}}$ at a given frequency $\omega_i$. The formulation of Eq.~\ref{eq:usual-spod} is convenient for homogeneous baseflows as $\bm{\mathsf W}$ is therefore defined in the space domain as a one-dimensional baseflow profile and projects on a single spanwise-wavenumber $\beta_j$ of the frequency vector $\hat{\bm{\mathsf X}}$. This structure allows to directly perform the product~\ref{eq:usual-spod} with a constant and real weight matrix $\bm{\mathsf W}$ in the \textit{space} domain. 

In light of this, reconsidering the decomposition of Eq.~\ref{eq:usual-spod} for a $N-$periodic baseflow requires to use the set of spanwise-wavenumber defined by $\gamma_M$ and its associated frequency vector $\hat{\bm{\mathsf X}}_{\omega_i,\gamma_M}$. Then, the weight matrix $\bm{\mathsf W}$ has to be defined for a transverse baseflow unit (a single streak pattern for instance). This leads to a dimension mismatch between $\hat{\bm{\mathsf X}}_{\omega_i,\gamma_M}$ being in defined in the frequency-wavenumber domain for $N_\gamma$ spanwise modes and $\bm{\mathsf W}$ in \textit{space} domain for $N_z$ spanwise points. A solution would be to perform the computation of Eq.~\ref{eq:usual-spod} in the frequency-space domain, but the cost compared to an homogenous flat plate would be multiplided by $N_z$ making the SPOD computation a tedious operation. 


Thus, an efficient formulation requires $\bm{\mathsf W}$ to be transformed to the spanwise-wavenumber domain and defined over the harmonic set of modes $\gamma_0$ to obtain $\hat{\bm{\mathsf W}}_{\gamma_0}$. Although, by property of the Fourier transform, a direct replacement of  $\bm{\mathsf W}$ by  $\hat{\bm{\mathsf W}}_{\gamma_0}$ and $\hat{\bm{\mathsf  X}}_{\omega_i,\beta_j}$ by $\hat{\bm{\mathsf X}}_{\omega_i,\gamma_M}$ in the inner product of Eq.~\ref{eq:usual-spod} is not mathematically equivalent to a formulation of Eq.~\ref{eq:usual-spod} in the frequency-wavenumber domain. Nevertheless, this can be solved by starting with the frequency-space domain formulation of Eq.~\ref{eq:usual-spod} for a frequency-space solution, with the inner product written as,
\begin{align}
    \left\langle \hat{\bm{\mathsf X}}_{\omega_i},\hat{\bm{\mathsf X}}_{\omega_i} \right\rangle_W &=  \hat{\bm{\mathsf X}}^H_{\omega_i} \left(\bm{\mathsf W} \hat{\bm{\mathsf X}}_{\omega_i}\right). \label{eq:inner-freq-space}
\end{align}
%
Considering the unitary nature of the Fourier transform, here in matrix notation, $\bm{\mathsf F}^H_\beta \bm{\mathsf F}_\beta=\bm{\mathsf I}$, and noting $(\tilde \bullet)$ a vector also Fourier-transformed to the wavenumber domain, Eq.~\ref{eq:inner-freq-space} can hence be expanded as follows,
\begin{align}
    \left\langle \tilde{\bm{\mathsf X}}_{\omega_i},\tilde{\bm{\mathsf X}}_{\omega_i} \right\rangle_W &= \hat{\bm{\mathsf X}}^H_{\omega_i} \bm{\mathsf F}^H_\beta \bm{\mathsf F}_\beta \left(\bm{\mathsf W} \hat{\bm{\mathsf X}}_{\omega_i}\right)  \nonumber \\
    &=  \tilde{\bm{\mathsf X}}^H \widetilde{\bm{\mathsf  W} \bm{\mathsf X}}, \nonumber \\
    &=  \tilde{\bm{\mathsf X}}_{\omega_j}^H \left(\tilde{\bm{\mathsf  W}} \circledast  \tilde{\bm{\mathsf X}}_{\omega_j}\right) . \label{eq:inner-floq}
\end{align}

The product $\widetilde{\bm{\mathsf  W} \bm{\mathsf X}}$ is then written as the convolution $(\circledast)$ of the Fourier-transformed vectors. This yields the equivalent formulation of the $\left\langle \hat{\bm{\mathsf X}}_{\omega_i},\hat{\bm{\mathsf X}}_{\omega_i} \right\rangle_W$  inner product in the frequency-wavenumber domain. 

Hence the CSD is computed for coherent structures of periodicity $M$ such as,
\begin{align}
    \bm{\mathsf S}_{\omega_i,\gamma_M} &= \tilde{\bm{\mathsf  X}}^H_{\omega_i,\gamma_M} \tilde{\bm{\mathsf W}}_{\gamma_0} \circledast \tilde{\bm{\mathsf  X}}_{\omega_i,\gamma_M}, \label{eq:freq-wave-spod}
\end{align}
where the set $\gamma_0$ is always used for the weight matrix as it is of harmonic periodicity by definition. 
From these derivations, the cost reduction of the SPOD formulation obtained with Eq.~\ref{eq:freq-wave-spod} can be estimated to be at least of a factor $N$. But, this reduction can be improved even further by considering that only a certain amount of wavenumbers $k$ in the sets $\gamma_M$ will have a relevant energy. Therefore, filtering out the wavenumbers of negligeable energy offer a further SPOD cost reduction, up to more than an order of magnitude compared to the initial formulation in frequency-space domain for this $N-$periodic baseflow. Additionally, this Floquet formulation improves the convergence of the SPOD modes compared to a direct fully 3D approach, as the $N-$periodic structure of the coherent structures is being directly embedded in the formulation of the eigenvalue decomposition and has not to be statiscally converged.


\bibliography{references} 

\end{document}